\documentclass[preprint]{aastex}

\usepackage{graphicx}
\usepackage{amsmath}
\usepackage[stable]{footmisc}

\newcommand{\ec}{$\eta$~Car}
\newcommand{\degree}{\ensuremath{^\circ}}

\begin{document}

\title{Critical Differences and Clues in Eta Car's 2009 Event\altaffilmark{*,**}}

\author{Andrea Mehner\altaffilmark{1}, 
        Kris Davidson\altaffilmark{1}, 
        John C.\ Martin\altaffilmark{2}, 
        Roberta M.\ Humphreys\altaffilmark{1}, 
        Kazunori Ishibashi\altaffilmark{3}, 
        Gary J.\ Ferland\altaffilmark{4}}

  \altaffiltext{*} {Based on observations obtained at the {\it Gemini\/} Observatory, which is operated by the
Association of Universities for Research in Astronomy, Inc., under a cooperative agreement
with the NSF on behalf of the {\it Gemini\/} partnership.}
  \altaffiltext{**} {Based on observations made with the NASA/ESA {\it Hubble Space Telescope}. STScI is operated by the association of Universities for Research in Astronomy, Inc. under the NASA contract  NAS 5-26555.} 
  \altaffiltext{1} {Department of Astronomy, University of Minnesota, 
       Minneapolis, MN 55455, USA}   
  \altaffiltext{2} {University of Illinois Springfield, Springfield, 
       IL 62703, USA}  
  \altaffiltext{3} {Department of Physics, Nagoya University, Nagoya 464-8602, Japan}
  \altaffiltext{4} {Department of Physics \& Astronomy, University of 
       Kentucky, Lexington, KY 40506, USA}  

\email{}

\begin{abstract}
We monitored Eta Carinae with {\it HST\/} WFPC2 and {\it Gemini\/} GMOS
throughout the 2009 spectroscopic event, which was expected to differ from
its predecessor in 2003 \citep{2005AJ....129..900D}.  Here we report major  
observed differences between events, and their implications.  Some of these 
results were quite unexpected.  
(1) The UV brightness minimum was much deeper in 2009.  This suggests that 
physical conditions in the early stages of an event depend on different 
parameters than the ``normal'' inter-event wind.  Extra mass ejection 
from the primary star is one possible cause.  
(2) The expected \ion{He}{2} $\lambda$4687 brightness maximum was followed 
several weeks later by another.  We explain why this fact, and the timing 
of the $\lambda$4687 maxima, strongly support a ``shock breakup'' hypothesis 
for X-ray and $\lambda$4687 behavior as proposed 5--10 years ago.  
(3) We observed a polar view of the star via light reflected by dust in 
the Homunculus nebula.  Surprisingly, at that location the variations 
of emission-line brightness and Doppler velocities closely resembled 
a direct view of the star;  which should not have been true for any 
phenomena related to the orbit. This result casts very serious doubt 
on all the proposed velocity interpretations that depend on the secondary 
star's orbital motion.  
(4) Latitude-dependent variations of \ion{H}{1}, \ion{He}{1} and
\ion{Fe}{2} features reveal aspects of wind behavior during the event. 
In addition, we discuss implications of the observations for several 
crucial unsolved problems.  
\end{abstract}


\section{Introduction}    

            Eta Carinae  is exhibit A for episodic mass 
loss near the top of the H-R diagram,  for the physics of giant 
eruptions (``supernova impostors'') and subsequent recovery,  the 
behavior of outflows above the Eddington limit, polar winds, 
several exotic nebular processes,  etc.   
Most of these topics remain poorly understood, but there is no reason 
to think that $\eta$ Car structurally differs from other extremely 
massive stars.  A likely companion object affects the phenomena,  
but does not by itself constitute an ``explanation.''

Beginning in the mid-1940s $\eta$ Car began to exhibit occasional 
spectroscopic changes that we now recognize as a  5.5-year 
spectroscopic/photometric cycle.   Occasionally its 
high-excitation 
\ion{He}{1}, [\ion{Ne}{3}], [\ion{Fe}{3}] emission lines disappear 
for a few weeks or months \citep{1953ApJ...118..234G,1984A&A...137...79Z} while other changes also occur, 
specifically in the X-ray (e.g., \citealt{1997Natur.390..587C,1999ASPC..179..266I,1999ApJ...524..983I}) and infrared flux (e.g., \citealt{1994MNRAS.270..364W,2001MNRAS.322..741F}).  These ``spectroscopic 
events'' recur with a 
period close to 2023 days \citep{1996ApJ...460L..49D,1994MNRAS.270..364W,1999ASPC..179..221D,2000ApJ...528L.101D,2004MNRAS.352..447W,2006ApJ...640..474M,2008MNRAS.384.1649D,2010NewA...15..108F}. 
They have been attributed to (1) eclipses of a hot secondary star by the 
primary wind \citep{1997NewA....2..107D,1999ASPC..179..266I,1999ASPC..179..295S,2002A&A...383..636P}; 
(2) disturbances in the primary wind triggered by a companion star near 
periastron \citep{1997NewA....2..387D,1999ASPC..179..304D,2003ApJ...586..432S,2006ApJ...640..474M};  (3) a thermal/rotational recovery cycle \citep{1984A&A...137...79Z,2000ApJ...530L.107D,2003ApJ...586..432S,2005ASPC..332..101D};  
or (4) a breakup/collapse of the wind-wind collision structure due to known 
shock instabilities \citep{2002ASPC..262..267D,2003ApJ...597..513S, 
2006ApJ...640..474M,2006ApJ...652.1563S}.  These ideas are not mutually 
exclusive.  Either (2) or (3) would be significant for massive-star 
physics in general, because they may require an undiagnosed instability   
near the Eddington limit.

Observations of the 2003.5 event appeared to favor possibility 2 and 
especially 4, but did not rule out number 3, and the likely geometry 
suggests that  an eclipse 
probably occurred with lesser consequences.   Photometric behavior, 
a chaotic X-ray behavior, and an unpredicted \ion{He}{2} $\lambda$4687 outburst 
\citep{2004ApJ...612L.133S,2006ApJ...640..474M} were especially significant.

Meanwhile {\it the longer-term behavior changed dramatically.\/}  
The central star brightened rapidly after 1998 
\citep{1999AJ....118.1777D,2004AJ....127.2352M,2006AJ....132.2717M,2010AJ....139.2056M},  
and major spectral features differed between the 1998.0 
and 2003.5 events \citep{2005AJ....129..900D}.  Destruction and/or 
lessened formation of dust played a role, but the root cause must involve 
a secular change in the UV flux or the wind density or both.   
Thus, from the viewpoint 
of 2007--2008, observations of the expected 2009.0 event merited a high 
priority for comparisons with 2003.5 and 1998.0.  Unfortunately the Space 
Telescope Imaging Spectrograph on the {\it Hubble Space Telescope\/} ({\it HST\/} STIS)
had failed in 2004, and thus was not available to separate the star from 
ejecta (see below).  Later, after the  event was observed as discussed in 
this paper, STIS became operational again and proved that 
emission lines from $\eta$ Car's wind had greatly weakened since 2004;  
the mass-loss rate had probably decreased by a factor of the order of 
2 or 3 in a time span of several years (\citealt{2010ApJ...717L..22M};  
see also \citealt{2009ApJ...701L..59K,2010ApJ...725.1528C}).  
Therefore the 2009.0 spectroscopic event occurred in physical 
circumstances appreciably different from its predecessors.
This should be helpful for deducing the nature of the event, on the 
same principle as varying parameters in an experiment.

Ground-based spectra of $\eta$ Car represent an unresolved mixture 
of the central star plus bright ejecta located at 
$r \sim$ 0\farcs2 to 2{\arcsec}. 
Fortunately, though, the central star has brightened more than the nearby 
ejecta in the past decade,  so ground-based  observations of it have 
become relatively less contaminated than at earlier times.\footnote{
    Here ``central star'' really means the opaque primary wind. 
    The secondary star is also included, but its spectrum is too 
    faint to be observable at accessible wavelengths. \label{foot:foot1}}      
Thus, in 2007 we began to observe the star and several offset positions 
in the Homunculus nebula with GMOS on the {\it Gemini-South\/} telescope. 
Apart from the question of secular changes, we planned to measure 
some aspects of the spectroscopic event better than had been 
done in 2003 and earlier.  A primary goal was to monitor the behavior 
of the peculiar \ion{He}{2} $\lambda$4687 line with more frequent observations 
during the 2009 event. We also observed the spectrum reflected by dust 
in the southeast (SE) Homunculus lobe, a ``polar'' view of the star.   
     
In this paper we discuss the resulting spectra and the light curve of 
the central star.  Some aspects of the 2009 event have been reported 
by other authors:  \citet{2010AJ....139.1534R} described 
the behavior of H$\alpha$, \citet{2009ApJ...701L..59K} and 
\citet{2010ApJ...725.1528C} commented on the X-rays, 
\citet{2010A&A...517A...9G} reported high-velocity material, \citet{2010NewA...15..108F} discussed photometry, and  
\citet{2011arXiv1104.2276T} have recently discussed the 
\ion{He}{2} $\lambda$4687 emission.  Here our scope is broader.  As  
we explain below, some of the differences compared to the 2003 event
provide unexpected new evidence for specific phenomena;  UV photometry 
of the central star indicates changed conditions;  reflected spectra 
showing the polar view  appear inconsistent with published 
models of the velocities; and we strongly disagree with 
Teodoro et al.\ concerning the past observational record 
and theoretical interpretations.   Our main new conclusions are 
that (1) differences between the 2003 and 2009 events give valuable 
and specific clues to the phenomena;  (2) a ``shock breakup'' scenario proposed 
a decade ago now seems almost inescapable; (3) the observed Doppler 
velocities are far less straightforward than most authors have assumed; 
and (4) realistic quantitative models -- as opposed to qualitative 
scenarios or idealized simulations --  are badly needed but will be 
very difficult to achieve.  We also comment on many other observational 
and theoretical factors.  Some important ideas that were discussed 
in connection with the 2003 event remain valid.

{\it HST\/} photometry and {\it Gemini\/} GMOS observations are described 
in the next section. The 2008--2010 light curve of the central star and 
the unusually deep minimum during the 2009 event are discussed in Section 3. 
In Section 4 we  discuss the peculiar \ion{He}{2} emission during the event, 
its connection with the X-ray flux curve, and the ``FOS4'' polar view 
of \ion{He}{2} emission compared with our direct line-of-sight view. 
The changing wind geometry during the 2009 event is discussed in Section 5.  
In the discussion section we summarize the results, emphasize the new 
information given by the 2009 event, and raise some outstanding questions.

\section{Observations and Data Reduction}  

\subsection{\textit{HST\/} Photometry with WFPC2 and STIS}
\label{sec:photometry}

We have monitored the brightness changes of the central star in several band-passes  with photometry from {\it HST\/} ACS/HRC and WFPC2 images and STIS spectra since 1998 \citep{2004AJ....127.2352M,2006AJ....132.2717M,2010AJ....139.2056M}. 
During the 2009 event we monitored the brightness of the central 
star with the {\it HST\/} WFPC2 camera using the F255W and F336W filters.   
F255W samples the NUV ``\ion{Fe}{2} forest'' \citep{1979A&A....71L...9C,1986ESASP.263..303A,1989ApJS...71..983V}, which greatly increases  
in opacity during a spectroscopic event \citep{1999ASPC..179..227D,2000AIPC..522..439G}.  
F336W includes the Balmer continuum 
augmented by various emission lines.  These filters have been calibrated 
for direct comparisons with the {\it HST\/} ACS/HRC F250W and F330W filters
\citep{2005PASP..117.1049S}.  Our own careful checks 
have led us to conclude that comparisons in these filters across 
instruments are valid for $\eta$ Car.
The images were reduced using the standard STScI data reduction 
pipeline. Calibrated  fluxes were measured in a  0\farcs3 diameter 
weighted aperture following procedures described in our previous papers 
which, combined with the spatial resolution of the {\it HST,\/}  minimize 
the influence from nearby bright ejecta. ACS-equivalent photometry 
was also synthesized 
from {\it HST\/} STIS data before mid-2004 and after mid-2009;  spectra 
were extracted with a weighted parabolic cross dispersion profile similar 
to the virtual aperture used to measure ACS/HRC images, convolved with 
the filter functions, and integrated \citep{2006AJ....132.2717M}.  The 
WFPC2 photometry from 2008.7 to 2009.3 and STIS synthetic photometry from 
2009.6  to 2010.6 is listed in Table \ref{tab:table1}.  Earlier data  from 
ACS/HRC and WFPC2 images and STIS spectra can be found in the papers 
cited above.

\subsection{\textit{Gemini\/} GMOS Observations}   
\label{sec:geminiobs}  

To cover the 2009.0 event, we obtained ground-based slit spectroscopy of 
$\eta$ Car  with the {\it Gemini-South\/} 
Multi-Object Spectrograph (GMOS) beginning in 2007 June through 2010 January.  
In most cases, we used the B1200 line grating at three tilt angles to 
cover the spectrum from $\lambda$3700 to $\lambda$7500 \AA. 
The 0\farcs5-wide slit, oriented with a position angle of 160\degree, 
was placed at four different positions: on the star, 
two offset positions $\pm$0\farcs75 relative to the star, and 
at a position known as ``FOS4,'' 4\farcs5 from the star in the SE lobe 
of the Homunculus.  (Operationally this was done by offsetting the slit  
$-2${\arcsec} parallel to itself.)  The star's polar spectrum is reflected 
by dust at FOS4.\footnote{   
      The name ``FOS4'' originated when it was a target for the 
      {\it HST\/} Faint Object Spectrograph in the 1990's    
      \citep{1995AJ....109.1784D,1999ASPC..179..107H,1999A&A...344..211Z}.   
      Apart from small pointing differences, FOS4 is the same as the 
      ``Center of SE Lobe'' in Figure 2c of Humphreys et al.\  and 
      Figure 3c of Zethson et al.  It was further discussed by 
      \citet{2003ApJ...586..432S}, \citet{2005AJ....129.1694W}, 
      and \citet{2005A&A...435..303S}.  The 1991--1997 {\it HST\/} FOS 
      data may be worth re-examining today for comparison with STIS 
      observations made in 1998--2010. \label{foot:foot2}}  

Since most of this paper is concerned with the \ion{He}{2} $\lambda$4687 
emission and features at nearby wavelengths, we concentrate 
on the blue spectra of the star and FOS4.  
Table \ref{tab:table2} is a journal of those observations. True slit 
positions vary slightly from the four locations described above, due 
to a combination of blind offset errors and differential 
atmospheric refraction, which were comparable. No corrector was available, and our observing goals usually did not 
allow observations with a vertical slit orientation (``the parallactic 
angle''). Therefore in each case we calculated the apparent position 
of the star as a function of wavelength, and used the offset slit 
position that was closest to the star for blue wavelengths.   
This procedure was adequate because, in effect, the slit positions 
overlapped due to the combination of slit width 
and seeing;  we did not attempt to measure accurate absolute fluxes.  
 For more information see Technical Memo 14  
on the $\eta$ Car Treasury Program website.\footnote{  
     http://etacar.umn.edu/treasury/techmemos/pdf/tmemo014.pdf \label{foot:foot3} }

We prepared two-dimensional spectrograms using the standard GMOS data reduction pipeline 
in the {\it Gemini\/} IRAF package, and extracted one-dimensional spectra via a routine 
developed earlier for use with {\it HST\/} STIS \citep{2006ApJ...640..474M}.
At each wavelength our software integrates the counts 
along a line perpendicular to the dispersion, weighted by a mesa-shaped 
function centered on the local spectral trace.  We used 
a mesa function with base-width = 11 pixels and top-width = 7 pixels,
about 0\farcs8 and 0\farcs5 respectively.  The seeing was 
roughly 0\farcs5--1\farcs5, so each GMOS spectrum discussed   
represents a region about 1{\arcsec} across.  The spectra were rectified using a LOESS fit \citep{LOESS1,LOESS2}. The pipeline wavelength 
calibration was improved using the interstellar \ion{Ca}{2} K absorption line 
at $\lambda$3935 \AA\ and the interstellar absorption line at 
$\lambda$5782 \AA. The absolute wavelength scale was obtained with 
{\it HST\/} STIS spectra that have better wavelength calibrations.  
Throughout this paper we quote vacuum wavelengths and heliocentric Doppler velocities.

\subsection{Concerning Times and Phases}  
\label{sec:timing} 

When referring to ``phase'' in the 5.5-year cycle, we consistently use the 
Treasury Project system with period 2023.0 days, which has been employed 
since 2003 without any need for revision; see http://etacar.umn.edu/, 
comments at the end of Section 2 in  \citealt{2010ApJ...710..729M}, and an Appendix to this paper.  In this 
paper, time within a spectroscopic event is denoted by $t$, such that 
$t = 0$ at MJD 54860.0 (2009 Jan 29), MJD 52837.0 (2003 Jul 17), etc.   
Periastron most likely occurs within a few days of $t = 0$, but that 
was not the reason for the choice of zero point.   
\citet{2011arXiv1104.2276T} arbitrarily use a different zero point that 
corresponds to $t \approx -18$ d, but the rationale for the long-standing 
Treasury Project system is noted in the Appendix.  


\section{Light Curve of the  Star 2008--2011} 
\label{sec:phot}

The light curve of the central star from {\it HST\/} data has two crucial 
advantages that have been overlooked in some discussions of $\eta$ Car's
photometric record. First, all ground-based photometry includes ejecta at 
radii $r > 0\farcs15$, practically unrelated to short-term variations of
the central star.  This contamination is remarkably strong, varies with
time differently than the star, and {\it is difficult to quantify.\/}
In recent years the ejecta accounted for 40--70\% of the brightness 
in the best ground-based photometry at visual wavelengths;    
see Figure 3 in \citet{2006AJ....132.2717M}.
Since this relative fraction has lately been decreasing on a timescale
of $\sim$ 10 years (see the ref.\ just cited), future ground-based photometry
may  eventually become representative of  the star but pre-2005 measurements 
were dominated by ejecta.  The contamination was probably still important 
during the 2009 event, but the amount is unclear;  
see \cite{2010ApJ...717L..22M} and the discussion below.
Fortunately the high spatial resolution of {\it HST\/} allows us to
sample the central object itself, $r < 0\farcs15$, with little extraneous
contamination -- albeit with less precision than high-quality 
ground-based photometry (cf.\ \citealt{2009A&A...493.1093F,2010NewA...15..108F} 
and http://etacar.fcaglp.unlp.edu.ar/).
A second obvious and important advantage of {\it HST\/} is access to the UV region
$\lambda < 300$ nm, where the largest photometric changes occur
during a spectroscopic event (see refs.\ in Section \ref{sec:photometry}).

\begin{figure}     
\epsscale{0.5}
\plotone{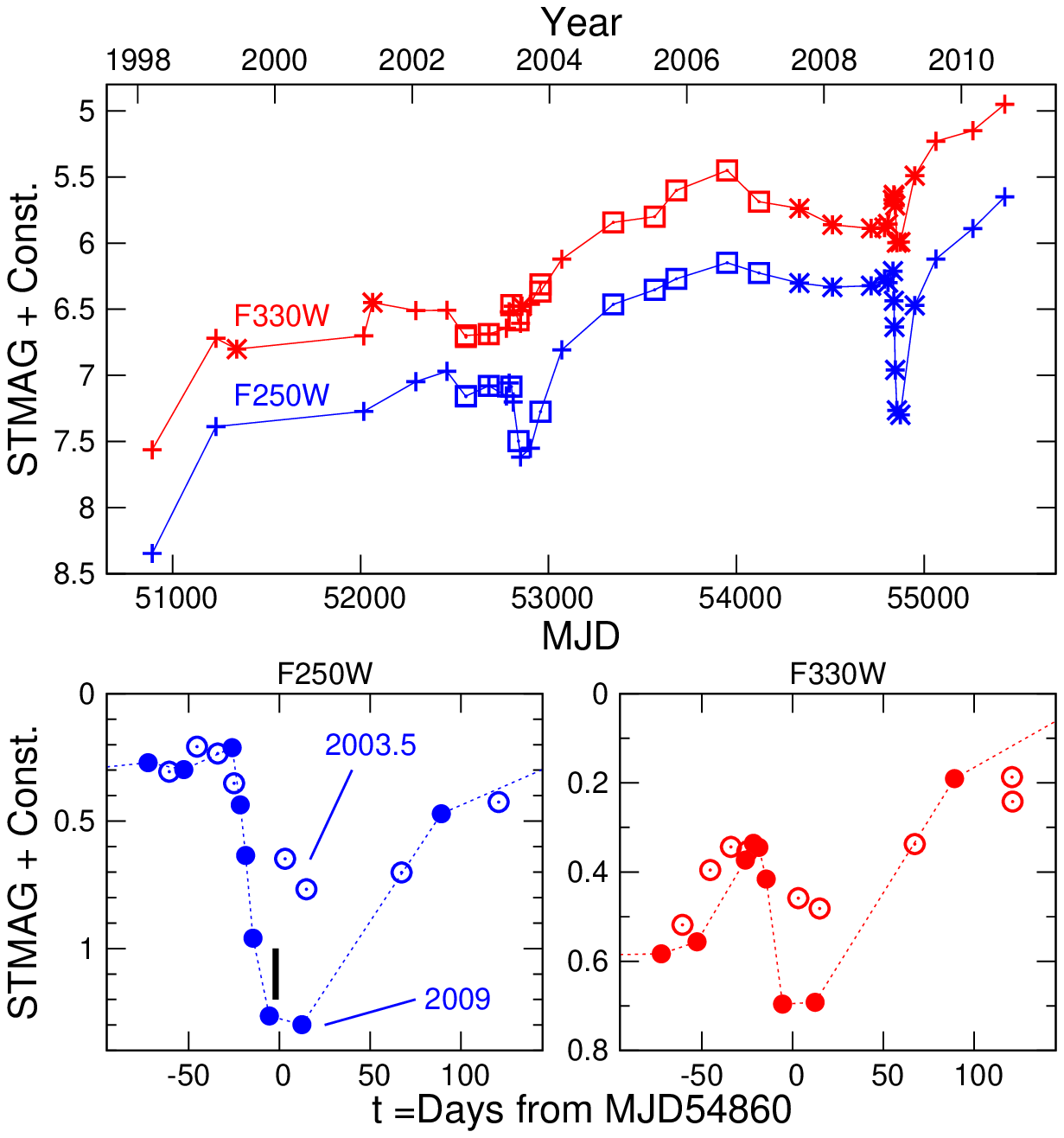}  
\caption{{\it HST\/} UV photometry. The upper panel shows the light curve from 1998 to 2011 (boxes: ACS/HRC, stars: WFPC2, crosses: STIS/CCD, values for F330W are shifted by -0.4 mag). The two 
lower panels compare  
 F250W and F330W brightness variations during the  
2009.0 spectroscopic event (filled circles and dotted line)  with the  
2003.5 spectroscopic event (open circles). Day ``0'' corresponds to
MJD 52837 for the 2003.5 event and MJD 54860 for the 2009.0 event. The black mark in the lower left panel indicates the date of the V band minimum observed by \citet{2010NewA...15..108F}. \label{fig:fig1}}
\end{figure}

Indeed the {\it HST\/} UV data illustrate an important point of this paper 
that is not obvious in the ground-based photometry:  {\it $\eta$
Car's 2009.0 event differed from its 2003.5 predecessor.\/}
Figure \ref{fig:fig1} shows the light curve in the F250W and F330W filters 
from 1998 through 2010 based on {\it HST\/} ACS/HRC and WFPC2 images and 
STIS spectra. The two smaller panels expand the light curve around the time 
of the event and show the minima for the 2009.0 and 
2003.5 events superimposed. The depth of the minimum in the 2009 event 
was about 1.1 mag in the F250W filter, which is sensitive to the 
\ion{Fe}{2} forest, and 0.4 mag in the F330W filter.  By contrast, the 
corresponding brightnesses decreased only half as much in 2003.  It is 
clear from the timing of the observations that this difference is not simply 
due to a missed minimum in 2003.5. In order to have a 1-mag amplitude 
undetected in the 2003.5 observations, the 250 nm brightness would need 
either a strong negative ``spike'' only a few days long, or an appreciably 
delayed minimum in terms of phase in the 5.5-year cycle, or both.  Either 
of these possibilities would be as remarkable and unexpected as a major 
difference in amplitudes.  (Unfortunately the minimum near 1998.0 is 
almost unknown.)

\citet{2010NewA...15..108F} observed a similar but much less dramatic 
difference between events at visual wavelengths. In 2008--2009 
they observed decreases of 0.15 to 0.25 mag in the UBVRI bands, 
only 0.02--0.03 mag deeper than during the 2003.5 event. 
It is unclear whether the difference in the depth of the minimum was 
small because
they observed longer wavelengths, or because they measured the 
brightness of star plus inner ejecta rather than the star alone.  
Unfortunately we could not obtain suitable {\it HST\/} data at blue 
to red wavelengths in 2005--2009, because even the shortest 
integration time allowed for WFPC2 would have greatly over-exposed 
the star with any of the filters appropriate for photometry at 
$\lambda > 360$ nm.

The ground-based visual light curves suggest a minimum in 2009 between 
$t = -2$ d and $+3$ d (\citealt{2010NewA...15..108F}, observations at 
MJD 54858--54863;  $\; t$ was defined in Section \ref{sec:timing} above).
In {\it HST\/} data, observations at $t = -5$ d and $+12$ d showed the lowest 
brightness, in agreement with their data.
Other aspects of the 2009 event are consistent with broad-band optical and 
infrared observations of past events 
\citep{2003IBVS.5477....1F,2003A&A...412L..25V,2004MNRAS.352..447W,2010NewA...15..108F}.
The star began brightening at an accelerated rate about 40 to 50 days prior 
to the onset of the event.  It brightened about 0.1 mag in F250W and 0.2 mag 
in F330W before it steeply declined.  Then, after about 150 days in F250W  
and about 100 days in F330W, the star returned to pre-event levels and
resumed its long-term brightening trend.

Why have the events become successively deeper in UV photometry 
(Figure \ref{fig:fig1})?  Here is one line of reasoning.  
The primary star's ill-defined UV photosphere is located in the opaque 
wind, far outside the star's surface.  If the wind density is 
gradually decreasing on a time scale of $\sim$ 10 years (Section 1 above,  
\citealt{2010ApJ...717L..22M}, \citealt{2006AJ....132.2717M}), 
then the photosphere seen at most times 
must be shrinking and becoming hotter.  During a spectroscopic 
event, however, the mass-loss rate may be enhanced, at least at low 
latitudes.   This possibility depends on the secondary star's 
tidal/radiative influence and on the size of the primary wind's 
acceleration zone \citep{1997NewA....2..387D,1999ASPC..179..304D,
2001ApJ...553..837H,2003ApJ...586..432S,2006ApJ...640..474M}.  
During an event the UV photosphere may therefore depend on 
different parameters than it does between events.  
(In principle, near periastron the secondary star might even be able 
to induce an outflow that is unrelated to the primary's usual wind.  
We do not advocate such an   extreme model here, but it illustrates 
the basic idea.)   Meanwhile the \ion{Fe}{2} forest around 
$\lambda \sim 260$ nm \citep{1999ASPC..179..227D,2001ApJ...553..837H} 
is sensitive to temperature and density.    For these reasons, one should 
not be surprised if the UV photospheric size, temperature, and brightness 
during events follow a different long-term trend than they do at other 
phases of the cycle (Figure \ref{fig:fig1}).  The same considerations 
apply to latitude and longitude dependences of the inner wind, even 
if there is no enhancement of total mass-loss rate 
\citep{2003ApJ...586..432S}.


\section{\ion{He}{2} $\lambda$4687 and X-rays}    
   \label{sec:heiixrays}

\ion{He}{2} $\lambda$4687 provides extraordinary clues to the nature 
of the spectroscopic event.  This is by far the highest-excitation 
feature known in $\eta$ Car's UV-to-IR wind spectrum, it appears only 
briefly at a certain stage in each event, and it probably signals 
a flood of very soft X-rays.  \citet{2004ApJ...612L.133S} first drew 
attention to it, but \citet{2006ApJ...640..474M} disagreed with their 
flux measurements and interpretation, and \citet{2011arXiv1104.2276T}
concur with the latter in most respects.  Based on the Martin et al.\ 
analysis, we can summarize the relevant physics:  
  \begin{itemize}  
  \item  Since the observed feature almost certainly results from 
     He$^{++}$ $\rightarrow$ He$^+$ recombination,   a temporary 
     source of He$^+$-ionizing photons ($h\nu > 54$ eV) is required. 
     Shocked gas flowing through the wind-wind collision zone does not 
     produce enough $\lambda$4687 emission via normal cooling.  
  \item  Nearly all authors agree that the two stars produce very 
     little radiation above 54 eV.  Therefore the 
     relevant photons are most likely 54--500 eV X-rays
     produced in the wind-wind shock structure.  Shocked gas of the 
     primary wind, with pre-shock velocities below 600 km s$^{-1}$, 
     is favorable for creating soft X-rays. 
  \item  The most suitable locale for $\lambda$4687 emission is in  
     the primary wind just before it encounters the colliding-wind shocks, 
     and/or in locally cooled condensations within the shocked region.  Since 
     the primary-wind shock is unstable 
     \citep{2002A&A...383..636P,2003ApJ...597..513S}, 
     these two choices may co-exist in roughly the same large-scale   
     volume if one smooths over the complex small-scale structures. 
     In either case $\lambda$4687 is excited via photoionization by 
     the soft X-rays mentioned above.   
  \item  The most plausible energy source is the primary wind. A naive 
     assessment predicts that the soft X-rays are inadequate for this 
     purpose by a factor of 3--10, but several effects improve the 
     efficiency. Martin et al.\ described radiative excitation effects 
     that amplify $\lambda$4687;  instabilities in the shocked region 
     tend to increase the number of very soft X-ray photons; and a brief  
     rise in the primary wind  outflow (hinted by other observations) 
     would also help.  With reasonable parameters, these details can 
     enhance the  \ion{He}{2} $\lambda$4687 flux by a sufficient factor. 
  \item  The supply of soft X-rays can temporarily rise to very high 
     levels if the fast secondary-wind shock becomes unstable like the 
     primary-wind side.   In that case the entire wind-wind interface 
     ``disintegrates'' and ``collapses,'' and a chaotic ensemble of 
     subshocks and oblique shocks may exist for a few days or weeks. This phenomenon may explain the brevity of the 
     $\lambda$4687 flash as well as the disappearance of 2--10 keV X-rays.     
  \end{itemize}

Some of these statements are controversial, but no other quantitative 
analysis has been published.  \citet{2004ApJ...612L.133S}  
proposed that \ion{He}{2} emission  occurs in the acceleration zone 
of the secondary wind, a much smaller region than those 
mentioned above.  If one employs consistent physical parameters, 
their model predicts a $\lambda$4687 flux two or three orders of 
magnitude too weak \citep{2006ApJ...640..474M,2006ApJ...652.1563S}.  
\citet{2006ApJ...652.1563S} also focused on the inner wind of the secondary 
star, but their scenario was very different, emphasizing a collapse of the 
shock structure followed by accretion onto the secondary.  They gave 
qualitative arguments for enhanced $\lambda$4687 emission in specified  
circumstances, but did not quantify the excitation physics.  
Their model includes some appealing components which we note in later 
sections below.  \citet{2011arXiv1104.2276T} recently indicated  
agreement with most of the above outline,  and did not attempt 
a new theoretical investigation.  In summary, the Martin et al.\ 
assessment is still the only detailed account of the radiative 
processes;  and so far there have been no strong arguments against it 
apart from geometrical details.  Below we shall mention various agreements 
between authors, and then some crucial disagreements.

\subsection{The first \ion{He}{2} Maximum during the 2009.0 Event\footnote{Throughout this section we use a time measurement $t$, with $t=0$ corresponding to MJD 54860, see Section \ref{sec:timing} and Appendix.\label{foot:foot4}}}
    \label{sec:episode1}    

\begin{figure}    
\epsscale{0.4}
\plotone{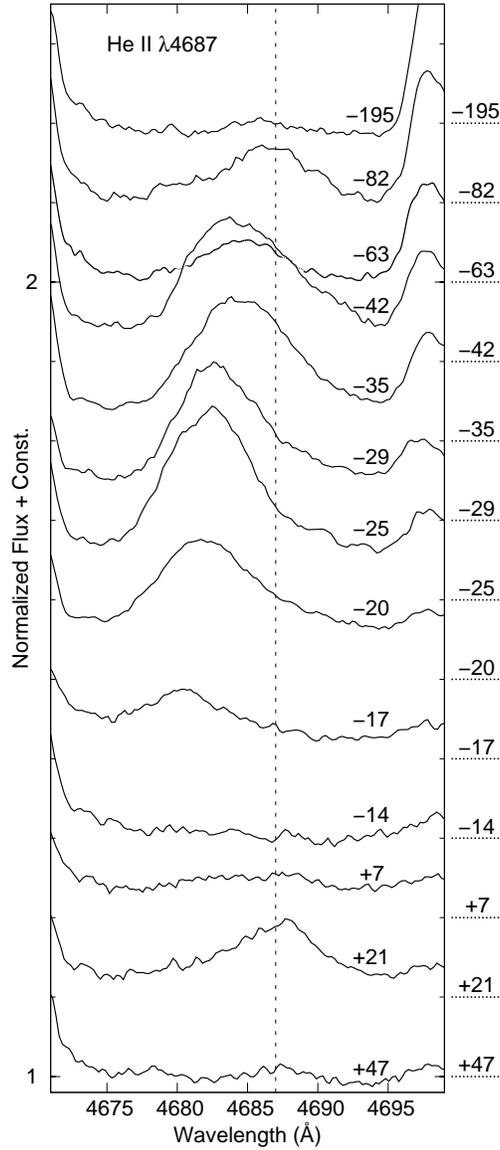}
\caption{Time sequence of the \ion{He}{2} $\lambda$4687 emission during the 2009 event in {\it Gemini\/} GMOS observations. Continuum was normalized to unity at $\lambda$4740 and is indicated with horizontal dotted lines on the right side. Offset between tracings is 0.1. Number of days before ($-$) and after (+) MJD 54860 are indicated next to each spectrum. The dotted vertical line indicates the position of \ion{He}{2} $\lambda$4687 at zero radial velocity. \label{fig:fig2}}
\end{figure}

Figure \ref{fig:fig2} shows a time sequence of the 
\ion{He}{2} $\lambda$4687 profile during the 2009.0 event, 
based on {\it Gemini\/} GMOS data.   One can 
liken its appearance to a wave that first moves leftward in the  
figure, and then is reflected back toward the right.  
(Note the reversed asymmetry of the profile at $t = +21$ d 
vs.\ $-20$ d.   Here ``reflection'' alludes to the line profile, 
not a real physical reflection.)     
The greatest source of uncertainty in this feature's strength is the 
underlying continuum level, which, following \citet{2006ApJ...640..474M}, 
we estimate by interpolation  between $\lambda$4605 and $\lambda$4744 
{\AA}.\footnote{   
   \citet{2004ApJ...612L.133S} greatly underestimated the peak $\lambda$4687 
   luminosity in 2003 because they chose too high a continuum level, shown 
   in Figure 3 in \citet{2006ApJ...640..474M}.  \citet{2011arXiv1104.2276T} 
   later adopted the Martin et al.\ method. \label{foot:foot5}}   
Observations reported by \citet{2011arXiv1104.2276T} agree well with ours.  
At its maximum, the \ion{He}{2} $\lambda$4687 emission extended across 
20 {\AA} or more (${\Delta}V > 1200$ km s$^{-1}$) and had a much larger 
flux than one might guess from the apparent size of the line profile.  
This is true because the apparent ``continuum'' on each side of the line 
includes considerable $\lambda$4687 emission;  see Figure 3 in Martin et al., 
Figure 2 in Teodoro et al., and the marks on the right side of our 
Figure \ref{fig:fig2}.    Part of the 20 {\AA} line width may conceivably 
be due to Thomson scattering rather than bulk velocities.  Here we shall 
ignore some minor anomalies,   e.g., the profile seen at $t = -35$ d was 
blueshifted less than either of its neighbors at $-42$ and $-29$ d.

\begin{figure}   
\centering
\epsscale{0.5}
\plotone{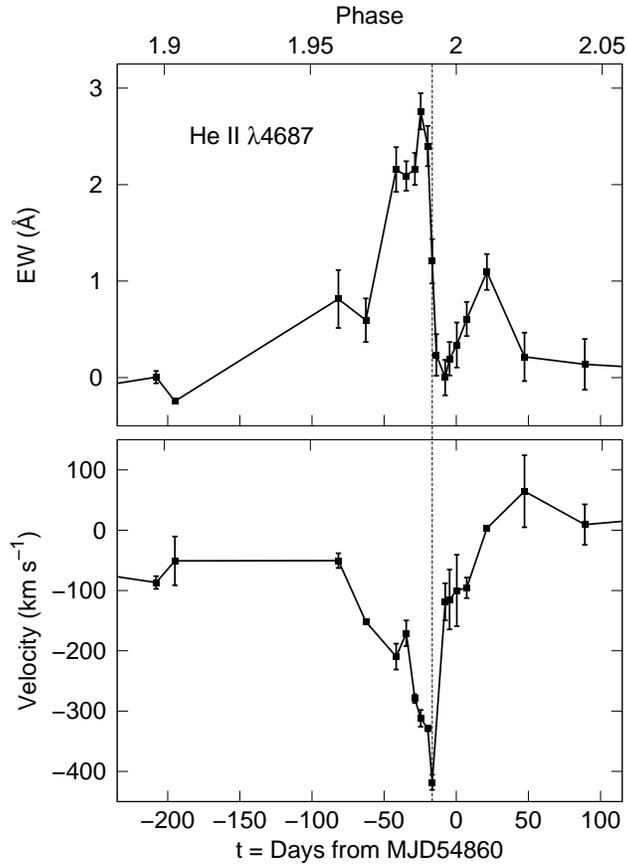}
\caption{Equivalent width and radial velocity measurements of the \ion{He}{2} $\lambda$4687 emission on the star during the 2009 event in {\it Gemini\/} GMOS data. The dotted vertical line indicates the time of maximum negative radial velocity which occurs at the flux-decline-midpoint. This is also true for previous events \citep{2004ApJ...612L.133S}. \label{fig:fig3}}
\end{figure}

Figure \ref{fig:fig3} shows the time development of equivalent width and 
Doppler velocity for \ion{He}{2} $\lambda$4687 in our {\it Gemini\/} data.   
Here ``equivalent width'' refers to flux between 4675 and 4694 {\AA} 
($-770$ to $+450$ km s$^{-1}$); other spectral features hide the 
farther line wings.  Our equivalent width measurements are listed in 
Table \ref{tab:table3}. 
``Velocity'' in Figure \ref{fig:fig3} refers to the line's peak.   
As Teodoro et al.\ have noted, the main,  
double-peaked, 40-day $\lambda$4687 flash in 2009 closely resembled the 
2003.5 event.   Its flux grew concurrently with the 
{\it decline\/} of 2--10 keV X-rays (Section \ref{sec:xrays} below);  the 
maximum equivalent width agreed with the Martin et al.\ value for 2003.5 
within measurement uncertainties; the time of significant brightness 
extended over 4--6 weeks; and the decline occurred in only one week.

The most precise time marker for $\lambda$4687 emission is the midpoint 
of its decline, which occurred at MJD 54843 or $t = -17$ d. 
The corresponding time in 2003 was close to MJD 52821 
\citep{2004ApJ...612L.133S},  and the difference of 2022 $\pm$ 2 d matches 
the consensus 2023-day spectroscopic period (see Appendix).   The 
decline-midpoint in our {\it HST\/} photometry at $\lambda \sim$ 250 nm 
occurred within a day or two of the same time (Section \ref{sec:phot}).
Moreover, within measurement errors, the most negative $\lambda$4687 
Doppler velocity also coincided with the flux-decline-midpoint.   
(Steiner and Damineli noticed the same coincidence in 2003, 
and considered it evidence for an eclipse.) At the time 
of maximum $\lambda$4687 brightness 8 days earlier, the line peak 
had $V_\mathrm{doppler} \approx -310$ km s$^{-1}$ (heliocentric);  
but then it rapidly moved to $-420$ km s$^{-1}$ at the decline midpoint.   
Later we shall indicate reasons why the similarity between the 2003 
and 2009 records is somewhat surprising.

\subsection{The ``Second Episode''}    
    \label{sec:episode2}    

A second, smaller \ion{He}{2} $\lambda$4687 flare occurred several  
weeks after the first,  around $t \sim +20$ d in Figures \ref{fig:fig2} 
and \ref{fig:fig3}.  Its significance will appear in 
Section \ref{sec:xrays} below;  it did not match 
the 2003 spectroscopic event.  Independent of that question, 
two aspects of the ``second \ion{He}{2} episode'' are noteworthy. 
First, the feature began to grow soon after its minimum 
at $t \sim -10$ d, and continued to do so for about a 
month.  Therefore the emitting region was not entirely eclipsed by the 
primary wind as advocated by \citet{2004ApJ...612L.133S}.  A second fact 
is the rapidity of change in Doppler velocity.  The first maximum or 
episode attained  $V_\mathrm{doppler} \, \sim \, -420$ km s$^{-1}$, 
but the emission that reappeared several weeks later had 
$V_\mathrm{doppler} \, \sim \, 0$ km s$^{-1}$.   Meanwhile the line profile's   
asymmetry reversed as mentioned earlier (Figure \ref{fig:fig2}).     
The overall velocity range greatly  exceeds the maximum 
projected {\it orbital\/} velocity variation for either star with 
any proposed  5.5-year orbit.  On the other hand, the line-of-sight   
wind velocities and post-shock velocities can easily differ by more 
than 400 km s$^{-1}$ at locations near the path followed by the secondary 
star near periastron.  This statement may have interesting and 
controversial implications for the orientation of the orbit, see 
Section \ref{sec:FOS4} and Section \ref{sec:discussion}.

Unlike the first \ion{He}{2} maximum, {\it the second episode in 2009  
apparently differed from the 2003.5 event.\/}   Here we disagree with 
Teodoro et al.\ about the 1992--2003 record, not 2009. 
The most reliable datum for this question was the {\it HST\/} STIS 
observation at MJD 52852 in 2003, carefully assessed by 
\citet{2006ApJ...640..474M}. Its timing corresponds to $t = +15$ d, 
only a few days before the phase of the second-episode peak seen in 
2009.   The 2003 STIS data showed no detectable line profile at that time,
see Figure 5 in Martin et al.   If one defines the emission feature as 
a bump or profile in the spectral tracing, then Martin et al.\ found a 
3$\sigma$ {\it upper limit\/} of only 0.15 {\AA} for the equivalent width 
at MJD 52852.  The points plotted in Figures \ref{fig:fig3}--\ref{fig:fig5}    
represent a different type of measurement, integrating a flux difference 
$f(\lambda) - f_{\mathrm{c}}(\lambda)$ over a 19 {\AA} interval.  Here 
$f_{\mathrm{c}}(\lambda)$,  the underlying continuum level, is the main 
source of error.  The 2003 
STIS data point at $t = +15$ d in Figure \ref{fig:fig4} may therefore be an 
overestimate, but the absence of a bump in the spectral tracing (see above) 
indicates 
that it is not likely to be an underestimate.  The observation is consistent 
with EW $\sim$ zero at that time, and excludes EW $\gtrsim 0.30$ {\AA};    
see Sections 5 and 6.3 in Martin et al.  
If the 2003 event had a second \ion{He}{2} episode matching the 2009 
case, then the equivalent width at MJD 52852 should have been 
close to 0.9 {\AA},  which would have been obvious in the STIS 
spectrum.  Conclusion:  {\it The 2003 event did not 
include a second \ion{He}{2} episode matching that seen in 2009.}  
If one did occur at the same phase, then in 2003 it must have been much 
weaker then 2009, even though the first \ion{He}{2} maximum was 
practically alike in both events.  But there is no good evidence for 
such an occurrence, and in Section \ref{sec:xrays} we favor the simple view 
that either 2003 had no second episode or else it was delayed.

The ground-based data are extremely unclear on this question,
especially since 
\citet{2011arXiv1104.2276T} disagree with \citet{2004ApJ...612L.133S} 
on major results even though they used similar, overlapping data sets.  
Only one data point is strongly pertinent:  Teodoro et al.\  
estimate EW($\lambda$4687) $\approx \, 0.65 \pm 0.16$ {\AA} at MJD 52841,  
eleven days before the 2003 STIS observation noted above.  In order to 
accept both this measurement and the STIS result,  one would need to 
postulate a sudden decrease between them.  This may be possible, but 
in that case the record would not resemble the 2009 second episode.  
A simpler explanation is 
that the MJD 52841 datum was an overestimate (see below).  Teodoro 
et al.\ also note a data point with EW $\approx 0.56$ {\AA} at MJD 48825 
in 1992 ($t \approx +34$ d), and perhaps two weaker ones, but the doubts 
expressed below apply even more forcefully to them.

Here we must digress on comparisons between ground-based data and 
{\it HST.}  All ground-based spectroscopy of $\eta$ Car is seriously 
(not just mildly) contaminated by emission-line ejecta at 
$r \sim$ 0\farcs2 to 2{\arcsec} \citep{1995AJ....109.1784D}.   
The size of this effect depends on the spectrograph aperture size and on 
variable atmospheric seeing.  Moreover, it was worse in the 1990's 
because the star was relatively faint then \citep{2006AJ....132.2717M}.  
There is concrete evidence that narrow emission lines in the ejecta 
tend to mimic a weak  ``\ion{He}{2} $\lambda$4687'' feature 
when the data are smoothed.  Details can be found in \citet{THESIS};   
here is a brief summary.   
{\it Gemini\/} data and slightly decentered STIS spectra show 
many weak narrow spikes near 4687 {\AA}.  Most of them are scarcely 
distinguishable from noise, except that their wavelengths recur 
in independent observations.  This fact is not surprising, since the 
ejecta are known to produce thousands of weak unidentified lines   
    \citep{2001PhDT.........1Z}.
When the spectra are heavily smoothed,  some of these tiny features 
blend to  produce a ``$\lambda$4687'' bump with height $\sim$ 0.5--1\% 
of the continuum level.  This bump consistently appears near the 
same wavelength, and it closely resembles the feature reported by 
\citet{2004ApJ...612L.133S} and \citet{2011arXiv1104.2276T} in data 
from Pico dos Dias {\it (OPD)\/} at times when $\lambda$4687 was weak.   
The {\it OPD\/} feature tends to be stronger by a factor of $\sim 2$, which can be 
ascribed to worse atmospheric seeing, a larger spectrograph aperture, 
and the fact that the apparent brightness ratio (ejecta/star) was much 
larger in the 1990's than it is today 
\citep{2006AJ....132.2717M,1995AJ....109.1784D}.   
(Despite this long-term trend, however, Teodoro et al.\ found the feature 
in data as late as 2010.) In summary, the {\it OPD\/}  data show symptoms of   
extraneous emission-line contamination, at a level that is not surprising 
but which strongly affects most of the $\lambda$4687 observations.   
{\it HST\/} STIS, by contrast, had good enough spatial resolution 
to separate most of the ejecta.   We conclude that   
pre-2004 ground-based data have very low weight compared to STIS 
in 2003 and many ground-based observations in 2009.  Consequently there 
is no significant evidence for a second \ion{He}{2} $\lambda$4687 flare 
around $t \sim +20$ d in 1992, 1998, and 2003.

There is an interesting puzzle in the observational record.  If we 
assume similarity among the 1992, 1998, and 2003 spectroscopic events, 
then \citet{2004ApJ...612L.133S} depicted a conspicuous second 
\ion{He}{2} episode {\it at a significantly delayed time $t$.\/}  
Their two relevant {\it OPD\/} observations occurred at $t \sim +30$ d 
in 1998 and $+50$ d in 1992, plotted in Figure \ref{fig:fig4} 
as green triangles.  These appeared to be stronger than the 
most doubtful cases mentioned above.  If valid, they imply second 
\ion{He}{2} flares in 1992 and 1998, but at phases roughly 30 days 
later than the 2009 case -- which would make sense for reasons 
noted in Section \ref{sec:xrays} below.   However, they contradict some 1992 
ESO data listed by \citet{2011arXiv1104.2276T}, and they 
are not mentioned in the latter paper even though it includes other 
data from the same instrument.  Either there were delayed second 
episodes in 1992 and/or 1998 (Steiner \& Damineli) or not (Teodoro et al.).
These inconsistencies reinforce our opinion that ground-based 
spectroscopy of $\eta$ Car in those years was far less robust than 
the 2003 {\it HST\/} results and various observations in 2009.

In summary:   One of the high-quality {\it HST\/} STIS observations    
in 2003 showed that the event in that year did not include a second 
\ion{He}{2}  episode matching the 2009 record;  and this assertion 
is not contradicted by any reliable ground-based data.  A weaker and/or 
later second episode may have occurred in 2003, as we discuss in 
Section \ref{sec:xrays}.


\subsection{\ion{He}{2}, X-rays, and Shock Breakup}  
\label{sec:xrays}

\citet{2006ApJ...640..474M} emphasized that \ion{He}{2} $\lambda$4687 
was anti-correlated with $\eta$ Car's observable 2--10 keV   
X-rays during the 2003.5 event;  the $\lambda$4687 flash occurred   
as the X-rays declined (Figure \ref{fig:fig4}).  
This fact appeared very consistent with a shock breakup as outlined 
below, and the 2009 ``second episode'' now strengthens the case.

\citet{2002ASPC..262..267D} remarked that known shock instabilities, 
rather than the  eclipse scenario that was popular at that time, can 
best  explain the rapid disappearance of $\eta$ Car's 2--10 keV 
X-rays during a spectroscopic event.  \citet{2003ApJ...597..513S} 
noted quantitative details, and \citet{2006ApJ...640..474M} 
emphasized the relevance of \ion{He}{2} $\lambda$4687 to this concept (see Section 9.2 of that paper).  
Various researchers later adopted essentially the same idea 
\citep{2008MNRAS.386.2330D,2009MNRAS.394.1758P,2011arXiv1104.2276T}.  
Two specific variants, with different causes but similar consequences, 
have been proposed.         
   \begin{enumerate}
   \item  A shock structure becomes unstable if radiative cooling 
   exceeds expansion cooling \citep{1992ApJ...386..265S}.  The slow 
   primary-wind shock of $\eta$ Car is very unstable in this regard, but 
   the faster secondary-wind shock stabilizes the overall structure in 
   calculated models \citep{2002A&A...383..636P,2003ApJ...597..513S}.  
   The secondary shock may become unstable {\it near periastron,} causing 
   the entire shock structure to disintegrate on a timescale of   
   10--30 days (Section 9.2 in \citealt{2006ApJ...640..474M}).  
   \item \citet{2006ApJ...652.1563S} drew attention to another  
   phenomenon that \citet{1990ApJ...365..321S} had studied for X-ray   
   binaries in general.  Near periastron, soft X-rays from the shocked 
   region can alter the ionization state of the secondary star's wind.  
   A higher degree of ionization tends to weaken the line-driven 
   acceleration,  resulting in a slower wind speed (``radiative 
   inhibition,'' a term later used by \citet{2009MNRAS.394.1758P}).  This in itself  
   would reduce the 2--10 keV flux;  but another consequence is that 
   the balance of wind momenta is altered, pushing the shocks closer 
   to the secondary star.  In an extreme case the primary wind can  
   entirely suppress the secondary wind.     
   \end{enumerate}  
Other instabilities certainly occur, such as Kelvin-Helmholtz which 
mixes gas from the two winds, and obvious thermal instability as shocked 
gas cools below $10^6$ K;  but the two processes listed above have mutually 
similar large-scale consequences and they might even develop together. 
Observable 2--10 keV X-rays rapidly and tremorously disappear as the 
highest temperature decreases;  a flood of soft X-rays is created by 
the chaotic ensemble of local shocks as the overall structure breaks 
up;  and the secondary wind may temporarily cease to exist.   
(Alternatively, it might survive in a slower form).   
Martin et al.\ noted that empirically, without referring to physics,
this view is supported by the growing unsteadiness of 
2--10 keV X-rays around their maximum.

Realistic simulation of these phenomena for $\eta$ Car  
would be extremely difficult,  and many sets of 
parameters must be explored.  Pending such models, one can estimate the 
relevant time scales from the overall size of the shocked region and 
local velocities within it:  say 1--3 AU and 100--800 km s$^{-1}$ 
respectively.  These values suggest ${\Delta}t \; \sim$ 2 to 50 days, 
consistent with the observed behavior.

Shock disintegration or collapse was originally only one among 
several competing explanations for the 2--10 keV X-ray crash during a 
spectroscopic event.  Then, in 2003 the previously unrecognized  
\ion{He}{2} $\lambda$4687 feature behaved 
just as the shock breakup idea would have predicted.  It 
appeared approximately when the hard X-rays peaked, it 
grew as they declined, and then ceased abruptly after a few 
weeks.  As \citet{2006ApJ...640..474M} noted, one would not 
expect this sequence and timing in an eclipse model like that 
of \citet{2002A&A...383..636P}, for example.

\citet{2011arXiv1104.2276T} have recently advocated a shock breakup  
model that closely resembles the earlier conclusions of Martin et al.  
We differ from Teodoro et al.\ in two important respects:  First, 
they state that \ion{He}{2} $\lambda$4687 originates ``4--5 AU downstream 
from the apex'' within the shocked gas.  However, as implied in point 3 
of the physics summary near the beginning of this section, there is 
no clear reason 
for this assumption.  Wherever very soft X-rays occur, they indirectly 
create $\lambda$4687 in the nearest suitable gas (which probably 
does not need to be shocked material).  Most of the applicable energy 
supply for soft X-rays near periastron is found at distances 
considerably less than 
4 AU from the pre-breakup apex or vertex.  Moreover, it seems 
possible or even likely that soft X-rays and $\lambda$4687 emission 
may continue for 5--20 days after the original large-scale shock structure 
has ceased to exist.  Our second difference is that Teodoro et al.\ 
give little attention to the evidentiary role of the unforeseen 
\ion{He}{2} second episode in 2009;  see below.

The 2009.0 event provides a valuable new clue, the \ion{He}{2} second 
episode discussed in Section \ref{sec:episode2}.  Figure \ref{fig:fig4} shows 
that its behavior was a reversal of the first episode seen 40 days 
earlier.   The $\lambda$4687 emission reappeared and peaked around 
$t \sim +20$ d while the 2--10 keV X-rays were still weak; 
then it declined concurrently with the growth of the X-rays.   
The overall time scale was comparable to that seen in the first episode.
In the shock-instability scenario, of course, we interpret this pattern 
as the re-formation of a large-scale shock structure when the relevant 
densities become sufficiently low for it to be quasi-stable.  This 
view raises several questions.  
\begin{itemize}  
  \item  As Figure \ref{fig:fig4} shows, the hard X-rays reappeared 
    earlier than expected in 2009, around $t \sim +30$ d rather than 
    $t \sim +60$ d as seen in 1998 and 2003 
    \citep{2009ApJ...701L..59K,2010ApJ...725.1528C}.  
    Given this difference, {\it should the earlier events have shown 
    the same \ion{He}{2} behavior pattern as 2009?}  The shock breakup hypothesis 
    ``predicts'' \ion{He}{2} $\lambda$4687 flares around $t \sim +50$ d 
    in 1998 and 2003, i.e., just before the hard X-rays reappeared.    
    Equally important, a $\lambda$4687 outburst as early as the 
    2009 case, $t \sim +20$ d, would not be expected.   
    The second \ion{He}{2} episodes would very likely have been 
    weaker in those earlier events, because the $\lambda$4687 
    production rate depends strongly on local densities.  Consider, 
    for example, a case with orbital eccentricity 0.9 and periastron 
    at $t = 0$, fairly consistent with most proposed orbit models.  
    Then the separation between stars was about 3.5 AU during 
    the pre-periastron $\lambda$4687 maximum ($t \sim -20$ d), but 
    more than 6 AU at $t \approx +50$ d just before the X-rays 
    reappeared in 1998 and 2003.  If we assume an unchanged primary 
    wind, these values imply a factor-of-three lower density at the 
    expected time of reappearance.  According to the Martin et al.\ 
    analysis, this would reduce the $\lambda$4687 flux by a factor 
    of 3--10 compared to its pre-periastron maximum -- depending on 
    local gas geometry, velocity gradients, etc.   Moreover, the  
    primary-wind density may have been enhanced in the weeks before 
    periastron \citep{1999ASPC..179..304D,2006ApJ...640..474M}. One should 
    therefore {\it expect\/} the second \ion{He}{2} episode to have 
    been weak in 1998 and 2003, possibly undetectable.  In 2009, on 
    the other hand, the second episode occurred earlier with higher 
    densities;  which is the main reason why it is a valuable clue.  
    (As we explain later, this statement includes a subtle complication 
    involving a secular decrease in wind densities.)  
  \item  {\it Is the observational record consistent with these 
    expectations?}  The answer is ``yes,'' but unfortunately the pre-2009 
    data were too sparse for confident conclusions.   We noted in 
    Section \ref{sec:episode2} that the second expectation, no substantial 
    $\lambda$4687 flare at $t \sim +20$ d in 2003, is confirmed by 
    STIS data.  The question of a later flare, however, amounts to a 
    conflict between observations reported by \citet{2004ApJ...612L.133S} 
    and those reviewed by \citet{2011arXiv1104.2276T}. The former 
    data strongly imply a $\lambda$4687 episode peaking around 
    $t \sim +50$ d in 1992 and 1998, see Figure \ref{fig:fig4} and 
    Section \ref{sec:episode2} above.  Teodoro et al., on the other hand, 
    omit those two data points and show no definite $\lambda$4687 
    emission after $t \sim +35$ d in 1992 and 1998, and no data points 
    in that time interval in 2003.  Neither 
    of these alternatives would contradict the shock breakup idea 
    (see above), but it would be useful to know whether or not a 
    detectable second  episode occurred.  Since the {\it OPD\/} data 
    are not public (unlike {\it HST\/} and {\it GMOS\/}), 
	it is difficult to assess these alternatives.   Meanwhile, the  
    2009 data and 2003 STIS data constitute the only satisfying results on this question. 
  \item   {\it Why did the \ion{He}{2} emission and hard X-rays reappear 
    earlier than expected in 2009?\/}  Referring only to the X-rays, 
    \citet{2009ApJ...701L..59K} proposed an explanation based on a decrease  
    in the primary star's wind.  They had a specific model in mind, but 
    the basic idea works for others as well.  Since 1999 there have been 
    hints that $\eta$ Car's wind density is (or was) becoming less dense   
    on a timescale of $\sim$ 10 years 
    (e.g., \citealt{2006AJ....132.2717M,2005AJ....129..900D}).    
    \citet{2010ApJ...717L..22M} recently reported strong spectroscopic 
    evidence for such a change.   Qualitatively, at least, one 
    expects the wind-wind shock structure to recover sooner after 
    periastron if the primary wind is less dense;  because in that 
    case the worst instabilities are weakened.  At first sight 
    the 2009 \ion{He}{2} second flare may seem paradoxical in 
    this view, because, as noted above, lower densities tend to 
    reduce the $\lambda$4687 emission efficiency.  A partial 
    explanation is that the shock-region density at the time 
    when it re-forms is not the same as the primary wind density 
    measured at some fixed radius;  the two stars were closer   
    together at $t \sim +20$ d in 2009 than they were at 
    $t \sim +50$ d in 2003.  This brings us to the next question 
    which also concerns timing and separation between the stars.  
  \item Suppose, as stated above, that the primary star's wind was 
    less dense in 2008--2009.  If so, {\it why didn't the shock structure 
    survive later than expected before periastron?}  With reduced 
    density $\rho_\textnormal{wind}(r)$, one would expect the shocks to 
    disintegrate at a smaller star-star separation, i.e., later. In fact   
    the 2009 X-ray crash and the main \ion{He}{2} $\lambda$4687 
    flash occurred almost exactly 2023 days after the 2003 
    event (Section \ref{sec:episode1} and Figure \ref{fig:fig4}).  
    Therefore, perhaps the characteristic densities at that time 
    did not directly represent the primary star's 
    ``normal'' steady mass-loss rate.  Instead, the inner 
    wind may have been tidally or radiatively enhanced during 
    the weeks before periastron, causing the relevant density 
    to be approximately the same for each spectroscopic event. 
    This hypothesis is not arbitrary or ad hoc, since it also 
    helps to explain a photometric puzzle mentioned at the end of 
    Section \ref{sec:phot} above, and it was suggested long ago 
    on other grounds (e.g., \citealt{1999ASPC..179..304D}, 
    \citealt{2006ApJ...640..474M}).   
  \end{itemize}

\citet{2011arXiv1104.2276T} have expressed agreement with most, but not 
all, of the points that we have quoted from \citet{2006ApJ...640..474M}. 
One disagreement pertains to the \ion{He}{2} ``second episode'' 
seen in 2009.  Above we emphasized the 
{\it anti-correlation\/} between $\lambda$4687 flux and hard X-rays 
at  critical times.  Teodoro et al., however, state instead that the two 
are {\it correlated\/} but $\lambda$4687 is delayed by 16.5 days.  
We strongly disagree, based on semi-theoretical reasoning plus an 
empirical fact.  (1) Those authors interpret the 16-day 
delay as the time required for shocked gas to flow from a favorable X-ray 
region near the vertex or apex of the umbrella-shaped shock structure, 
to another location where it has cooled enough to produce \ion{He}{2} 
emission.   For this explanation one needs a localized density 
enhancement, which must pass near the shock vertex and later reaches 
some particular radius at a well-defined time after flowing outward 
within the shocked zones. But these are unlikely assumptions in 
the context of an extended, chaotic, unstable shock structure.  There 
is no clear reason for the localized density enhancement;  only a tiny 
fraction of material flowing outward from either star passes close 
to the shock vertex;  gas is vigorously mixed and diffused within 
the shock structure;  the overall geometry is probably unstable 
at the critical time as noted above;  and there is no reason to assume 
that \ion{He}{2} emission occurs only beyond a certain distance from 
the vertex.  In other words the proposed rationale is very unclear. 
(2) Our second, more empirical reason is simpler:  
The 16-day delay proposed by Teodoro  
et al.\ is obviously inconsistent with the $\lambda$4687 second episode 
observed in 2009 (Figure \ref{fig:fig4}).  Maximum \ion{He}{2} emission 
occurred {\it after\/} the hard X-ray peak in  the first episode, 
but {\it before\/} the X-ray recovery in the second.

In summary:  Strictly speaking we cannot prove the shock breakup scenario, 
but the combined $\lambda$4687 and X-ray data are impressively consistent 
with it.  In terms of logic, the role of \ion{He}{2} $\lambda$4687 
has been as follows.  First, this feature 
was not recognized when the breakup idea was first proposed;   
but then it turned out to fit into that scenario in a remarkably 
natural way.  Later, when a post-periastron ``second \ion{He}{2} episode'' 
appeared in 2009 as discussed above, it too matched quite naturally, and 
the observed secular decrease in wind density provides a good 
explanation for the differences between 2003 and 2009.  In neither case 
was there any need to alter the concept to fit new observations.

In terms of theoretical development, \citet{2002ASPC..262..267D} and  
\citet{2003ApJ...597..513S} stressed that shock breakup is a 
reasonable concept for $\eta$ Car;  then \citet{2006ApJ...640..474M}     
and \citet{2006ApJ...652.1563S} argued that it is a very likely one.  
\citet{2009MNRAS.394.1758P} later cited other X-ray clues, notably 
that gas dynamic computations have not sustained the competing 
eclipse scenario.  Given these facts, and lacking a viable alternative 
explanation, we conclude that $\eta$ Car's colliding-wind 
shock structure does, indeed, disintegrate and collapse during  
a spectroscopic event.  Two separate questions are (1) whether  the 
primary star has a mass-loss outburst at about the same time, and (2) 
whether the secondary star accretes material then.  See 
Section {\ref{sec:discussion}} below.

Is \ion{He}{2} $\lambda$4687 emission detectable in $\eta$ Car only 
during spectroscopic events?  According to \citet{2004ApJ...612L.133S} 
and \citet{2011arXiv1104.2276T}, this feature has been present at  
other times.  On the other hand, \citet{2006ApJ...640..474M} 
found no broad \ion{He}{2} in non-event {\it HST\/} STIS spectra 
of the central star.   STIS mapping data in 2009 June and December  
show no broad $\lambda$4687 emission in extended regions around the star.  
Analyzing observations made with {\it HST\/} STIS, {\it Gemini\/} GMOS, 
and {\it Ir\'{e}n\'{e}e  du Pont\/} B \& C, 
we find that a weak emission feature exists
but it is only about 3 {\AA} wide, i.e., narrower than the \ion{He}{2} 
emission seen when it is bright. As noted in Section \ref{sec:episode2}, 
it is most likely a blend of very weak, narrow, unidentified lines 
and not broad \ion{He}{2} emission \citep{THESIS}.

\subsection{\ion{He}{2} $\lambda$4687 in the Reflected Spectrum from the Pole (FOS4)}
\label{sec:FOS4}    

Spectra reflected by dust in the Homunculus nebula give surprising 
new information about velocities in the 2009 event.  The known geometry 
of the bipolar Homunculus allows us to correlate each position in the 
southeast lobe with stellar latitude, assuming that the polar 
axis of \ec\ is aligned with the 
Homunculus axis.  FOS4 is a location near the center of the SE lobe 
(the one nearer to us)  which reflects a nearly pole-on view of the 
stellar wind, $\sim 75$\degree\ latitude  
\citep{2003ApJ...586..432S,1995AJ....109.1784D,1999A&A...344..211Z}.  
Our direct view probably represents $\sim 45$\degree\ latitude 
\citep{2001AJ....121.1569D,2006ApJ...644.1151S}.   
FOS4 is also useful for another reason, namely that ground-based spectra 
there are less contaminated by nebular lines than direct observations 
of the star are.  (The reasons for this fact are not entirely clear, but 
generally speaking our direct view of the star appears to have more 
circumstellar extinction than the average line of sight.)  FOS4 is  
located about 3\farcs7 south and 2\farcs5--3\farcs5 east of the 
central star.

A spectrum reflected in the Homunculus has a light-travel-time delay 
${\Delta}t$ corresponding to the additional path length, and an extra Doppler 
shift ${\Delta}V$ due to the ``moving mirror'' effect 
\citep{1987A&A...181..333M,2005A&A...435..303S}.  With standard assumptions 
about the Homunculus,  these are related by  
${\Delta}V/c = {\Delta}t/$(age), where ``age'' means elapsed time since the 
reflecting material was ejected in the 1840's. The value of ${\Delta}V$  
depends on location in a straightforward, observable way that depends on 
the shape of the Homunculus lobe, see Figure 4 in \citet{2001AJ....121.1569D}.  
In fact we can define FOS4 as the location where ${\Delta}V = +100$ km s$^{-1}$, 
which implies ${\Delta}t \approx 20$ d for the 2003 and 2009 spectroscopic 
events.   \citet{2005A&A...435..303S}, however, reported a time-delay of 
only 10 days for appearance and disappearance of \ion{He}{2} $\lambda$4687 
in 2003, comparing {\it VLT} UVES observations of FOS4 to observations by 
\citet{2004ApJ...612L.133S} which were centered on the star. If correct, 
this would cast doubt on standard analyses of the reflection model, 
Homunculus expansion, etc.  The good  time coverage of our 
{\it Gemini\/} GMOS data during the 2009 event, both on the star 
and on FOS4, allows us to re-evaluate this result. 
Figure \ref{fig:fig5} shows the equivalent width and radial velocity 
measurements of \ion{He}{2} $\lambda$4687 in spectra of the star in 
direct view and reflected at FOS4.  It also includes re-measured 
equivalent widths in UVES data during the 2003.5 event, shifted by 2023 days. 
The GMOS data show that the time-delay in the observed equivalent widths 
as well as radial velocities, between the direct view on the star and the 
pole-on view at FOS4, is about 18 days which confirms the expected 
${\Delta}t \; \approx 20$ d within the attainable accuracy. Also, 
re-measurement of the UVES data during 
the 2003.5 event according to the method employed by 
\citet{2006ApJ...640..474M} show that the earlier data are in accordance 
with the GMOS 18 days value.  We detected a positional gradient 
of ${\Delta}t$, which agreed with the simple model within the 
measurement errors.  The geometry of the reflection process 
therefore appears satisfactory.

Unexpectedly, we find that {\it the behavioral pattern of
\ion{He}{2} $\lambda$4687 emission is very similar when viewed from 
different directions,\/} i.e. in direct view of the star and reflected at 
FOS4, when we take the time-delay into account. Values for the equivalent 
widths and radial velocities are slightly smaller at FOS4 than in the 
direct view.  Figure  \ref{fig:fig5} shows these results, with Doppler 
velocities corrected for the moving-mirror effect ${\Delta}V$.      
\ion{He}{1} $\lambda$4714 seems to exhibit similar behavior, but cannot 
be measured well because its line profile changes shape in a more complex 
way than \ion{He}{2} $\lambda$4687 (Section \ref{sec:wind} below, and \citealt{THESIS}).     
{\it These results suggest that velocities of helium lines are not simply 
related to orbital motion of the secondary star,}             
if we assume that the orbit inclination is $i \sim$ 40--45\degree\ 
like the Homunculus midplane, see refs.\ cited above. 
In standard models the view from FOS4 should be almost perpendicular 
to the plane of the orbit, and therefore no large radial velocity 
variations should be observed there.  Thus the data at FOS4 are very 
surprising.

As examples, let us mention just two of the published models.     
\citet{2006ApJ...652.1563S} proposed that helium lines originate in the 
acceleration zone of the secondary star. Assuming we view the orbit at an 
inclination of 40--45\degree, then the \ion{He}{2} Doppler velocity variation 
observed in our direct view, $\sim 400$ km s$^{-1}$, implies values of  
$\sim 550$ km s$^{-1}$ in the plane of the orbit.  The variation 
at FOS4 is $\sim 250$ km s$^{-1}$;  if we interpret this too as 
the projection of a velocity in the orbital plane (70--80\degree\ inclination 
at FOS4), then true values of $\sim$ 750--1500 km s$^{-1}$ are needed.  
In order to attain high enough projected flow velocities, the Soker and 
Behar model would therefore require FOS4 to ``see'' 
\ion{He}{2} $\lambda$4687 emission formed farther out in the acceleration 
zone than the emission in our direct view.  This is not easy to arrange, 
because the emission process is inherently isotropic.  Even more 
significant, one would also need to explain velocity {\it variations\/} 
that greatly exceed any credible orbital velocity.   We shall comment 
later on possible alterations of the assumed inclination.  
Similar arguments apply to models in which the helium emission 
originates in or near the wind-wind collision region;  for instance 
the \ion{He}{1} interpretation by \citet{2007ApJ...660..669N} is quite 
inconsistent with the FOS4 data.  No matter whether one ascribes the 
observed Doppler variations to orbital velocities or to successive 
illumination of wind regions near the moving secondary star, FOS4 was 
expected to differ from a direct view of the star.  The differences 
should have been far more conspicuous than our observations show.

Given the consistent time-delay at FOS4, it is very difficult 
to imagine that the above result is mistaken;  but here we can 
offer only conjectures to explain it.       
One might reconcile the data with a different inclination $i$, 
such that the projected orbit appears alike from both our point of 
view and FOS4.  But that idea would ``multiply the hypotheses'' 
by requiring {\it two\/} unlikely assumptions:  (1) The orbit 
plane must be tilted 20--25\degree\ from the Homunculus midplane, 
and (2) the azimuthal direction of the tilt must be aligned with 
both our line of sight and that of FOS4, so they share nearly the same 
projected velocities in phase with each other.  In other words, 
both the latitude and longitude of the tilt must be suitable.    
This recourse seems too artificial and {\it ad hoc\/} 
to be an appealing first choice.  A better explanation 
might be that the observed Doppler variations represent   
``global'' changes in outward velocities, roughly spherical outflows 
that are not given a strong directionality by the secondary star.  
(An extreme form of this view might even 
return to the single-star mass ejection proposed by 
\citealt{1984A&A...137...79Z}.  We do not intend to go that far, but the 
question is worth contemplating.)  
A shock breakup model may conceivably act in a quasi-spherical way, 
with chaotic random velocity components during the critical 
time.  The same statement applies to enhanced mass-loss from the 
primary star.  Independent of these thoughts, {\it our FOS4 
results cast doubt on attempts to derive orbit parameters from 
apparent emission-line velocities.\/}


\section{The Changing Wind Structure during the 2009 Event}
\label{sec:wind}    

{\it HST\/} STIS data revealed that some spectral features depend on viewing direction and that the global stellar wind geometry changes during the cycle \citep{2003ApJ...586..432S}. The most dramatic effects occur at low latitudes, while the dense polar wind remains relatively undisturbed during an event. Departures from spherical symmetry are critical for theories of winds and 
instabilities in the most massive stars and we therefore re-examine selected 
spectral features at differing latitudes in our {\it Gemini\/} GMOS data.  
Smith et al.\ analyzed only three epochs after the 
1998 event; 1998 March (phase = 0.04), 1999 February (phase = 0.21), and 
2000 March (phase = 0.40). The improved time-sampling of the GMOS observations 
makes it possible to monitor changes before, during, and after the 2009 event.
Our results do not all agree with Smith et al., and some of the differences 
may signal real changes.

\paragraph{Hydrogen:}

H$\alpha$ and H$\beta$ emission lines are so bright in \ec\ that all H$\alpha$ and many of the H$\beta$ observations centered on the star were  overexposed in {\it Gemini\/} GMOS observations.\footnote{
{\it Gemini\/} GMOS does not support exposure times below 1 s. \label{foot:foot6}
} 
For this object H$\gamma$ is usually contaminated by other emission lines, so we  
analyzed H$\delta$.   \citet{2010AJ....139.1534R} have described the 
behavior of H$\alpha$ in 2008--2009, based on a large number of observations 
with higher spectral resolution;  but H$\alpha$ samples different 
properties of the system because it originates at much larger 
radii than H$\delta$ \citep{2001ApJ...553..837H}.

Figure \ref{fig:fig7} exhibits variations observed in the 
H$\delta$ P Cyg profile at several latitudes. This figure shows 
spectral tracings seen at the star and four locations to the southeast, 
at several phases close to the 2009 event. GMOS observations before the 
event, from 2007 June to the beginning of 2009 January, revealed no 
significant changes in the H$\delta$ P Cyg profiles.  At those times 
($t = -353$ and $-82$ d), substantial  P Cyg absorption was observed 
only at higher latitudes.  This fact is usually considered evidence that 
the density and/or ionization structure of \ec's current stellar wind 
outside an event is nonspherical \citep{2003ApJ...586..432S}.
For most of \ec's spectroscopic cycle, wind densities are expected to 
be higher near the poles, in accordance with theories of equatorial 
gravity darkening in massive rotating stars 
\citep{2000A&A...361..159M,2001A&A...372L...9M,2005ASPC..332..169O}.
But \citet{2010AJ....139.1534R} have noted an alternative explanation, in which the secondary star prevents hydrogen P Cyg absorption in the primary wind.  
A basic obstacle to settling this question is that the fractional 
population of H$^{0}$ in the $n = 2$ level, essential for Balmer 
absorption, is an intricate theoretical question that has not yet 
been explored for $\eta$ Car's wind.

Standard theory predicts higher wind velocities at high latitudes, 
and \citet{2003ApJ...586..432S} reported such a result extending almost 
to  $\sim -1000$ km s$^{-1}$ in the 2000 March STIS data (see their 
Figure 5).  However, we find $v_{\infty} \sim$ 500--550 km s$^{-1}$ 
for H$\delta$ absorption at all latitudes during \ec's normal 
state (Figure \ref{fig:fig7}).  One reason for this discrepancy may be 
that we use a different method 
to align the spectra.  Smith et al.\ corrected for ${\Delta}V$ in the 
expanding nebula (Section \ref{sec:FOS4} above) by aligning the blue side   
of H$\alpha$;  but that reference point in the line profile is itself 
affected by details of the P Cyg absorption.  We use, instead, several 
forbidden lines that are known to originate in the Weigelt knots with 
constant velocities much smaller than the discrepancy in 
question.  Also, the velocity structure of the wind may change 
throughout the cycle;  Smith et al.\ used observations at phase 0.40 
while our relevant observations are at phase 1.83 and 1.96.  
Our GMOS observation at phase 2.17 cannot be used to investigate this 
issue, since the H$\delta$ profile had obviously not returned to its normal state.
H$\delta$ should be practically as good as H$\alpha$ and H$\beta$ for 
this purpose, since its absorption component is stronger, relative 
to emission, than in H$\alpha$.  A real change in the velocities and 
latitude structure may have occurred between 2000 and 2008.

As already discussed by \citet{2003ApJ...586..432S} and 
\citet{2005A&A...435..303S}, the strong latitude dependence of 
Balmer P Cyg profiles does not apply during the events. In only a few 
days, between $t = -25$ d and $-20$ d, the P Cyg absorption at 
lower latitudes appeared and strengthened to the same depth as at 
higher latitudes. Strong absorption at all latitudes was observed 
until $t \approx +47$ d, i.e. for at least 70 days.  Observations 
from $t = +89$ d to $+344$ d showed only weak absorption at low latitudes 
while high latitudes continued to have strong P Cyg profiles;  
the system had almost returned to its pre-event state.

P Cyg absorption is always present in the higher Balmer lines at all 
latitudes, consistent with their formation regions closer to the star 
\citep{2005AJ....129.1694W}.  Still, during the events, their absorption 
deepened at lower but not higher latitudes.  

Doppler velocities of Balmer lines are difficult to assess because 
they may include two different but unresolved parts.  As noted in 
\citet{2011arXiv1104.1829M}, the main velocity of H$\delta$ emission remains fairly steady, 
while a second component appears to vary like \ion{He}{1} 
(see below). 
The two components overlap so much that 
neither can be studied individually.

\paragraph{\ion{He}{1}:}

Both the emission and the absorption components of \ion{He}{1} vary in 
strength and radial velocity throughout the cycle, see Figure \ref{fig:fig6} 
for measurements on the star. The equivalent width of the \ion{He}{1} 
emission was mostly constant during the cycle, increased before the 2009 
event, and then dropped into a temporary minimum. However, the emission 
line never disappeared entirely as would have been expected if the 
event were a true eclipse.  The emission lines shifted monotonically 
blueward throughout the cycle, terminating with an abrupt, large velocity 
shift of over $-100$ km  s$^{-1}$ to velocities of about $-250$ km s$^{-1}$ 
near the event, followed by a sharp rise to almost zero radial velocity 
-- very similar to the \ion{He}{2} $\lambda$4687 behavior noted in
Section \ref{sec:heiixrays} above.       
The radial velocity of the absorption lines showed a similar pattern with 
velocity shifts between $-300$ km s$^{-1}$ and  $-600$ km s$^{-1}$.\footnote{
   \citet{2007ApJ...660..669N} found similar values during the 2003.5 event. \label{foot:foot7}}
\citet{2003ApJ...586..432S} found that while the \ion{He}{1} emission faded 
at low latitudes during the 1998 event, the emission was relatively 
undisturbed at higher latitudes. However, GMOS observations show that the 
equivalent width of the \ion{He}{1} emission at FOS4 during the 2009 event 
followed basically the same patterns as the emission directly on 
the star. 

Figure \ref{fig:fig8} shows tracings of \ion{He}{1} $\lambda$4714 in GMOS 
observations at several latitudes close to the 2009 event. Outside the 
event,  from 2007 June to 2008 July, \ion{He}{1} lines had strong P Cyg 
absorption in spectra at low latitudes and only very weak P Cyg absorption 
in spectra at higher latitudes (see also \citealt{2003ApJ...586..432S}). 
Because of their limited time-sampling, Smith et al.\ were not able to 
observe changes in the \ion{He}{1} P Cyg absorption. This led them to 
conclude that for most of the cycle \ion{He}{1} absorption is present on 
the star but not at higher latitudes.  Their Figure 18 implies that the 
absorption at low latitudes would disappear during the event, though this 
is not explicitly stated in the paper.    However, in our data we find that 
shortly before the 2009 event, the \ion{He}{1} absorption increased at 
higher latitudes. At $t = -82$ d the P Cyg absorption was strong at all 
latitudes. Then, over the next 2 months the absorption weakened at all latitudes. 
Between $t \approx -5$ d and $t \approx +89$ d almost no absorption was observed.
By $t \approx +176$ d the \ion{He}{1} profile had returned to its normal state displaying strong absorption at lower and almost no absorption at higher latitudes.
Note that changes in the \ion{He}{1} P Cyg absorption occurred at least 
2 months earlier than those in the \ion{H}{1} P Cyg absorption, and 
the return to the pre-event state took up to 3 months longer for 
\ion{He}{1}.  
The presence of \ion{He}{1} P Cyg absorption at all latitudes shortly 
before the event may have implications for the shell ejection model favored 
by \citet{2003ApJ...586..432S}, since this observation is not directly 
accounted for by that model.

The geometrical volume filled by ionized helium around $\eta$ Car is 
highly relevant,  since \ion{He}{1} absorption as well as emission 
lines depend on He$^+$ $\rightarrow$ He$^0$ recombination.      
The zones of interest are photoionized chiefly by the hot secondary 
star, see \citet{2008AJ....135.1249H}, \citet{2010ApJ...710..729M},
and refs.\ therein.   If the primary wind between events has 
become less dense in the past few years \citep{2010ApJ...717L..22M}, then 
the He$^{+}$ zone photoionized by the secondary star should now occupy 
a much larger volume than it did at earlier times.    
A crude assessment of the helium-ionizing photon supply suggests that 
the He$^+$ zone most likely now wraps around the inner primary wind 
at times other than spectroscopic events.  In other words, the  
pseudo-hyperboloidal He$^+$/He$^0$ ionization front in the primary wind 
may now be concave toward the primary star. If so, then most lines of 
sight to the primary star -- including that for FOS4 -- must pass 
through some He$^+$ in the primary wind, unlike the case 10 years ago.  
Detailed work will be presented in a future paper.

\paragraph{\ion{Fe}{2}:}

Narrow lines and broad stellar wind emission of \ion{Fe}{2} are strong 
in direct spectra of \ec, but much fainter in the reflected spectrum at FOS4. 
As already noted by \citet{2003ApJ...586..432S}, the \ion{Fe}{2} lines 
resemble Balmer lines in that the \ion{Fe}{2} P Cyg absorption increases 
with increasing latitude and that the emission is weaker at higher latitudes.  
Figure \ref{fig:fig9} shows broad stellar wind emission of \ion{Fe}{2} in 
spectra at FOS4.  Spectra before $t \approx -82$ d showed only very weak 
absorption at FOS4, with maximum strength at  $\sim -$ 400--450 km s$^{-1}$, 
but the absorption feature then deepened and stayed strong until 
$t \approx +176$ d, i.e.\ for about 250 days. The absorption was strongest 
around $t \approx +10$ d.  The deepening of \ion{Fe}{2} absorption at FOS4 
was also observed during the 2003.5 event by \citet{2005A&A...435..303S}. 
Other species, such as \ion{Cr}{2}, \ion{Mg}{1}, and \ion{Ti}{2}, also 
developed absorption lines.

\citet{2003ApJ...586..432S} argued that \ion{Fe}{2} lines are formed in the same outer wind regions as \ion{H}{1} because these two species show similar 
latitude dependence and are likely to be coupled by charge exchange. However, their similarity is only true outside an event. During or close to an event the lines behave very differently; in contrast to \ion{H}{1}, \ion{Fe}{2} does not develop absorption in direct view.  (\ion{H}{1} lines develop strong P Cyg absorption at lower latitudes while the pole remains almost unchanged.)
Theoretically, \ion{H}{1} Balmer absorption is not really like most \ion{Fe}{2} 
absorption, since the hydrogen $n = 2$ level has a much higher energy than 
most \ion{Fe}{2} lower levels (10 eV compared to 0--3 eV).  The observed Balmer absorption lines 
may involve H$^+$ $\rightarrow$ H$^0$ recombination, analogous to 
\ion{He}{1} noted above.


\section{Critical Differences in the 2009 Event -- A Summary, Conclusions, and Problems}
\label{sec:discussion}   

\subsection{Primary Results}    

We monitored $\eta$ Car with {\it HST\/} WFPC2 and {\it Gemini\/} GMOS 
throughout its 2009 spectroscopic event.  Good time coverage with 
the GMOS slit oriented at a constant position angle made it possible to 
monitor spectroscopic changes as seen from a range of directions, 
some of them via reflected light.  In this paper we have described 
several important differences compared to previous events, some of them 
quite unexpected.  These results lend strong support 
to one important concept, the idea of shock breakup and collapse 
near periastron.   Regarding other aspects of the problem, the new 
information helps to exclude some proposed models, while -- as usual for 
this topic -- it also deepens some of the puzzles.

{\it HST\/} WFPC2 observations show that the minimum in the UV was much 
deeper for the 2009 event than for the 2003.5 event (Section \ref{sec:phot}).  
One possible explanation involves a mass-loss outburst or similar 
phenomenon, discussed in Section \ref{sec:unsolved} below.

Contrary to expectations, we find that the behavior of the \ion{He}{2} 
equivalent width and radial velocity reflected at polar location FOS4 
are very similar to direct observations of the star.  For this purpose 
an observed light-travel-time delay ${\Delta}t \approx 18$ d and 
moving-mirror redshift ${\Delta}V \approx +100$ km s$^{-1}$ must be taken 
into account.  Since the observed ${\Delta}t$ agrees well with the 
predicted value, it confirms that FOS4 really does ``see'' the star 
from a polar direction.  {\it The observed radial velocity behavior 
at FOS4 is a surprise,\/} because, contrary to proposed models, it 
fails to show major differences compared to our direct view of the star. 
This statement includes other effects as well as the orbital velocity.

We found that the ``second \ion{He}{2} $\lambda$4687 episode'' 
in 2009 was strongly anti-correlated with the X-rays, like a time-reversal 
of the main $\lambda$4687 outburst seen in 2003 and 2009.  
In Section \ref{sec:xrays} 
we argued that this result strongly supports the shock structure  
breakup/disintegration/collapse hypothesis, first proposed a decade ago 
as an explanation for X-ray behavior during a spectroscopic event.    
Strictly speaking this idea has not been {\it proven,\/} but now the 
following facts together have made it the most probable model.  
  \begin{enumerate} 
  \item  Likely physical conditions near periastron appear suitable for 
     the relevant instabilities 
     \citep{2003ApJ...597..513S,2006ApJ...640..474M,2006ApJ...652.1563S}.  
  \item  The \ion{He}{2} $\lambda$4687 outburst was not yet recognized 
     when shock breakup was first proposed in 2001--2003, see Section \ref{sec:xrays}
     above.  But after it was discovered and measured, this phenomenon 
     turned out to be very well suited to the idea.  
  \item  The 2009 event featured a new development, the second 
     $\lambda$4687 episode about 30 days after the first. As 
     we discussed in Section \ref{sec:xrays}, this was beautifully     
     accordant with the earlier-than-expected reappearance 
     of 2--10 keV X-rays.   
  \item  The time-scales are reasonable as noted earlier. 
  \item  From a theoretical viewpoint, other proposed explanations 
    (especially eclipses, which most authors favored until two or 
    three years ago) have much worse difficulties 
    \citep{2006ApJ...640..474M,2009MNRAS.394.1758P}.  
   \end{enumerate}  
But shock breakup is still a hypothesis, albeit a very strong one, not 
a quantitative theory.  Soft X-ray production during the collapse,  
and structural recovery later, have scarcely been investigated yet.   
A broad range of parameter space must be explored. 

Let us emphasize that shock breakup would explain only some aspects of 
a spectroscopic event.  Other phenomena, more fundamental for the basic 
physics of $\eta$ Car, are probably required in order to explain 
the spectroscopic and photometric changes.   Some of them are noted 
in Section \ref{sec:unsolved} below.

Results on spectral features of hydrogen, \ion{He}{1}, and other species  
(Section \ref{sec:wind}) are too diverse to summarize briefly.
Hydrogen P Cyg profiles at different latitudes throughout the cycle behave as already discussed in \citet{2003ApJ...586..432S}. Outside the  2009 event, 
\ion{H}{1} P Cyg absorption is observed at higher latitudes but not at lower, while during the event \ion{H}{1} P Cyg absorption is also strong at lower latitudes. GMOS data show that the P Cyg absorption at lower latitudes appeared suddenly within only a few days and was present for at least 70 days. We do not find a higher terminal velocity at higher latitudes as found by \citet{2003ApJ...586..432S}. The exact reasons have to be addressed in the future.

Helium P Cyg profiles showed an additional incident not discussed by 
Smith et al.  Those authors found that outside the events \ion{He}{1} P Cyg 
absorption is present at low latitudes but absent at higher latitudes, while during the events the absorption disappears at low latitudes, too. Our GMOS data show that shortly before the 2009 event \ion{He}{1} absorption increased at higher latitudes to similar strength as at low latitudes. Then the absorption decreased slowly at all latitudes. Changes in the \ion{He}{1} lines were observed already 2 months before the \ion{H}{1} lines showed any changes and they returned to their normal ``no-event''  state up to 3 months later than the \ion{H}{1} lines. 

\ion{Fe}{2} absorption, only present at higher latitudes, became very strong during the 2009 event for several months and behaved different than the \ion{H}{1} lines.

Further analysis is required with regard to the cause of those observed latitude dependent changes throughout the events. Is a minor shell ejection sufficient to explain them? Or do changing ionization fronts in the primary wind caused by influences of the secondary star and/or the moving wind-wind shock structure play a role?

\subsection{Unsolved Problems}   
\label{sec:unsolved}

A number of essential questions have not yet been answered even after 
years of observation and discussion.  Each requires theoretical work that no 
one has attempted at a realistic level of detail, and our 2007--2010 
results are pertinent to some of them.

\paragraph{Does the secondary star accrete material during a spectroscopic 
event?\/}  This possibility has been emphasized especially by Soker 
and colleagues (e.g. \citealt{2003ApJ...597..513S,2006ApJ...652.1563S,2007ApJ...661..482S,2009ApJ...701L..59K}).  
As we noted in Section \ref{sec:xrays}, near periastron the primary wind 
may entirely suppress the secondary wind, thereby allowing accretion 
onto the secondary star.  This possibility can be separated from other 
aspects of those authors' model.  It is a very appealing idea because 
it may explain a long-standing paradox concerning $\eta$ Car's \ion{He}{1} 
emission lines \citep{1999ASPC..179..304D,2008AJ....135.1249H}.   
These features depend chiefly on photoionization by the hot secondary 
star, they weaken during a spectroscopic event, and {\it they were not 
present before 1941.\/}  Merely saying that ``the secondary 
star was engulfed in dense gas'' does not constitute an explanation,  
since helium-ionizing photons  ($h{\nu} > 24.6$ eV) inevitably 
generate \ion{He}{1} recombination lines in the primary wind even 
at densities far above normal.  Roughly speaking, the brightness 
of \ion{He}{1} emission measures the far-UV photon supply.\footnote{  
   \citet{2010ApJ...710..729M} estimated $T_\mathrm{eff} \approx 40,000$ K 
   and $10^5 \, L_\odot \; \lesssim \; L \; \lesssim \; 10^6 \, L_\odot$ 
   for $\eta$ Car's secondary star.  We estimate that such a star 
   does indeed produce enough helium-ionizing photons to account for 
   the observed equivalent widths of $\eta$ Car's \ion{He}{1} lines. 
   In addition to the reasoning used for nebulae \citep{2006agna.book.....O}, 
   one must include a special enhancement factor  explained in Section 6 
   of \citet{2008AJ....135.1249H}.      \label{foot:foot8}
      }  
However, as \citet{2007ApJ...661..482S} noted, accretion can lower the 
secondary star's effective temperature by slightly expanding its photosphere.  
Even a modest reduction in temperature substantially cuts the output of 
helium-ionizing photons (\citealt{2010ApJ...710..729M} and refs.\ therein).  
Humphreys et al.\  remarked that $\eta$ Car's mass-loss rate may have been 
far above $10^{-3} \; M_{\odot}$ yr$^{-1}$ in the years 1900--1940;  the 
primary star's wind density near the secondary star then may {\it usually} 
have been as large as it is today near periastron -- i.e., very likely 
enough to crush the secondary wind. (See also \citealt{2009MNRAS.397.1426K}, 
concerning $\eta$ Car's Great Eruption.)

Accretion thus provides a satisfying explanation to the \ion{He}{1} puzzle; 
but is it correct?  Helium emission lines did not disappear in 2009 
(Figure \ref{fig:fig6}), but on the other hand this was presumably the 
least dense spectroscopic event on record.   Truly realistic simulations 
of three-dimensional accretion and resulting photospheric temperature 
will be extremely difficult, but they may prove necessary.  At present 
we cannot be certain that the secondary wind is suppressed near periastron.

\paragraph{Does each spectroscopic event include a mass-loss outburst by 
the primary star?}  This suggestion arose long ago because eclipse models 
appeared inadequate \citep{1999ASPC..179..304D}, and because 
\citet{1984A&A...137...79Z} had discussed the same idea in a single-star 
context. Today the  shock breakup concept probably explains the X-ray and 
\ion{He}{2} $\lambda$4687 behavior, but other spectral 
features involve larger spatial regions in the unshocked primary wind.  
Throughout this paper we have mentioned indications that the primary wind may have been 
disturbed.  Most of them have been noted before, and none is entirely 
satisfying, but their combination is highly suggestive:          
  \begin{itemize}  
  \item  Our UV photometry showed much deeper minima in 2009 than in 
    2003.  In Section \ref{sec:phot} we noted why this may be evidence 
    for extra material in the system during the early stages of 
    the event.  Qualitatively, a higher density due to temporary 
    causes would affect the \ion{Fe}{2} forest and UV photometry 
    in the desired direction.  Shock breakup, by contrast, does 
    not provide such an obvious explanation.  A similar argument 
    can be made regarding near-infrared free-free emission. 
  \item  Temporarily enhanced gas densities would help to destabilize 
    the shock structure, while also supplying extra energy for 
    \ion{He}{2} $\lambda$4687, as \citet{2006ApJ...640..474M} explained.
  \item  In Section \ref{sec:xrays} we noted that the 2008--2009 X-ray crash
    ``should have'' occurred later than predicted, since it is now known that the primary wind 
    had greatly decreased since 2003 \citep{2010ApJ...717L..22M,2009ApJ...701L..59K,2010ApJ...725.1528C}.  
    Extra ejected material dependent on other factors, however, would have 
    made the secular density trend temporarily irrelevant. 
  \item  In Section \ref{sec:FOS4} we stated that a reflected polar view 
     of the 2009 event closely resembled the behavior seen directly, 
     including the Doppler variations.  Any effect involving orbital 
     motion should have appeared different between the viewpoints.  
     A conceivable explanation is that velocities during the event
     may have represented instead a varying quasi-spherical outward 
     flow from the primary star.  
  \item Complicated changes in P Cyg absorption features,   
     described in Section \ref{sec:wind} above, suggest that column 
     densities increase especially at low latitudes during an 
     event (compare \citealt{2003ApJ...586..432S}). 
  \item  If the Eddington limit is taken into account (and, perhaps, 
     rotation), tidal forces are very likely strong enough near periastron to 
     alter the primary wind-acceleration zone \citep{1997NewA....2..387D}.  
     Various instabilities may have broadened the flow. 
  \end{itemize} 
Of course we do not claim that a mass ejection event would explain 
everything.  But it would help with all the above points, 
and there is no clear evidence against it.

The outburst conjecture has both milder and stronger variants.  For 
instance, the total mass-loss rate might remain constant during the event, 
but its latitude dependence briefly changes, causing densities to increase 
at low latitudes \citep{2003ApJ...586..432S}.   
At the other extreme, the base of the wind might be affected, not just 
its acceleration zone.  That case would signal an undiagnosed surface 
instability in the star itself.  How can this question be investigated?   
As a beginning, one might explore theoretically whether the photometric 
and spectroscopic observations can be explained by shock breakup 
{\it without\/} a primary wind outburst or disturbance.

The primary wind density has been decreasing and may continue to do so 
until the star has a normal line-driven wind \citep{2010ApJ...717L..22M}.  
If so, then the wind-wind shocks will very likely become stable even at 
periastron, ending the series of X-ray and \ion{He}{2} ``events'' 
as we know them today.   This might even occur within the next two 
or three 5.5-year cycles.  If, on the other hand, the secondary star 
is capable of triggering a tidal/radiative outburst near periastron 
(as discussed above), then the X-ray events will continue to occur.  
In that case the X-ray luminosity will be diminished, but not as severely 
as the wind density.  (This statement is based on simple considerations 
of the wind-wind momentum balance and cooling rate.) 

\paragraph{How can the orbital parameters be estimated?}  
The FOS4 results bolster our opinion that observed Doppler velocities 
cannot be used for this purpose until they are much better understood.  
There have been attempts to derive orbit parameters by directly 
identifying $\eta$ Car's observed Doppler variations with orbital 
velocities, like a classical spectroscopic binary (e.g., 
\citealt{1997NewA....2..107D,2007ApJ...660..669N,2008MNRAS.390.1751K}).  
This approach is questionable because neither the emitting nor the 
absorbing gas is expected to share the motion of either star;  efforts 
of this type have contradicted each other;  and now our FOS4 results 
create even greater doubts.   The observed velocities are more likely 
to represent samples of wind regions near the moving secondary star, 
but in this case one needs to model the complicated varying 
three-dimensional spatial volumes, local absorption may vary, and 
the FOS4 puzzle still applies.  Thus, until more is known about 
the projected velocities, orbit parameters must be based on other 
considerations.

\begin{figure}  
\epsscale{0.5}
\plotone{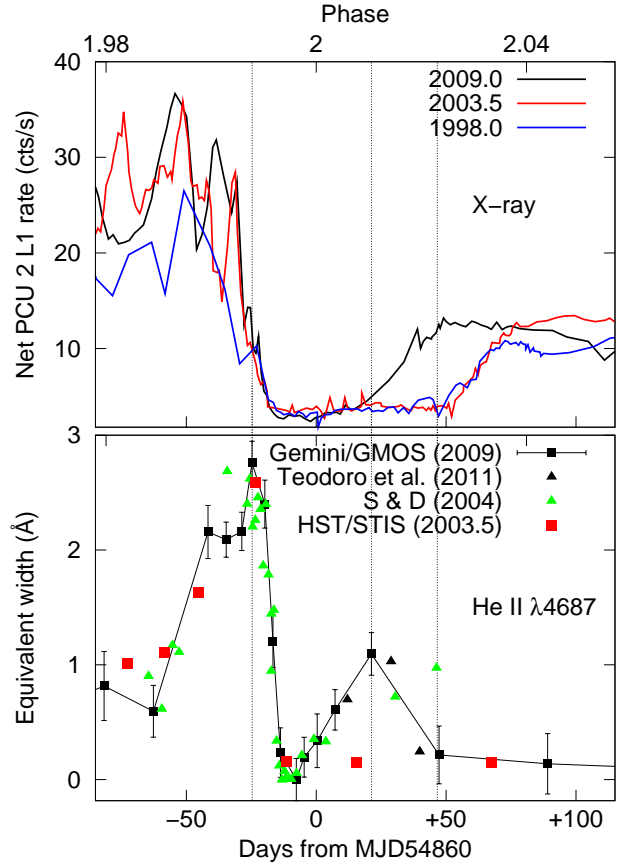}
\caption{TOP: X-ray light curve during the 1998, 2003.5, and 2009 events (from http://asd.gsfc.nasa.gov/Michael.Corcoran/eta\_car/etacar\_rxte\_lightcurve/index.html). BOTTOM:  Equivalent width of the \ion{He}{2} $\lambda$4687 emission during the 2009 event with {\it Gemini\/} GMOS (black squares) and selected data points (black triangles) from \citet{2011arXiv1104.2276T} which constrain the timing of the second episode during the 2009 event further. 
The red squares are measurements of {\it HST\/} STIS data during the 2003.5 event. No significant \ion{He}{2} emission was found in 2003 at $t \approx +20$ d in STIS data and measurements by \citealt{2004ApJ...612L.133S} during the 1992.5 and 1998 events (green triangles; these values are scaled by a factor of 3, see Footnote \ref{foot:foot5}) indicate a delayed second episode. This casts doubt on the timing of the second episode found by \citet{2011arXiv1104.2276T}. Values for older events are shifted by multiples of 2023 days. \label{fig:fig4}}
\end{figure}

\begin{figure}    
\epsscale{0.5}
\plotone{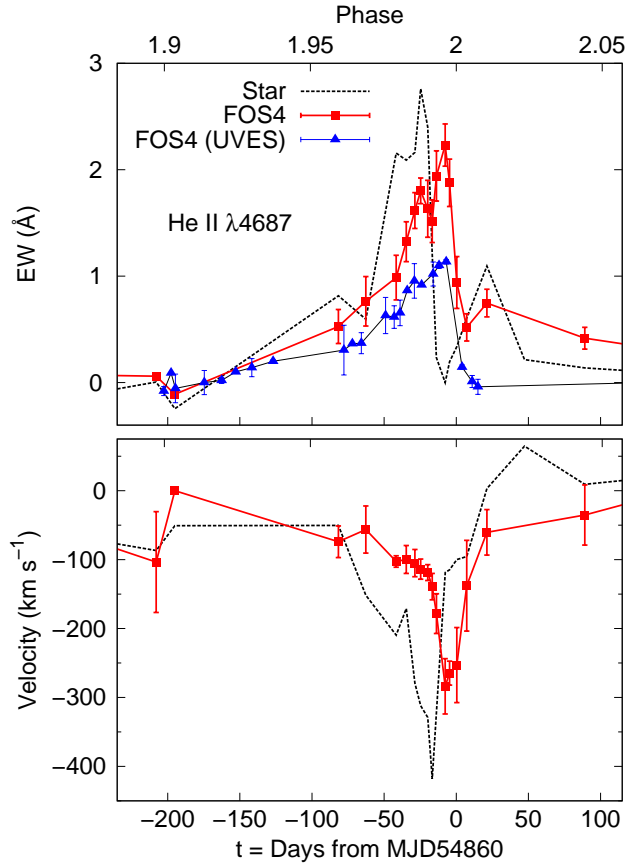}
\caption{Equivalent width and radial velocity of \ion{He}{2} $\lambda$4687 at FOS4 during the 2009 event observed with {\it Gemini\/} GMOS (red squares). Dashed curves are measurements on the star in direct view. The time-delay between spectra of the star in direct view and spectra at FOS4 is about 18 days. Re-measurements of {\it VLT\/} UVES data during the 2003.5 event are consistent with this finding (blue triangles, values are shifted by 2023 days).\label{fig:fig5}}
\end{figure}

\begin{figure}   
\centering
\epsscale{0.5}
\plotone{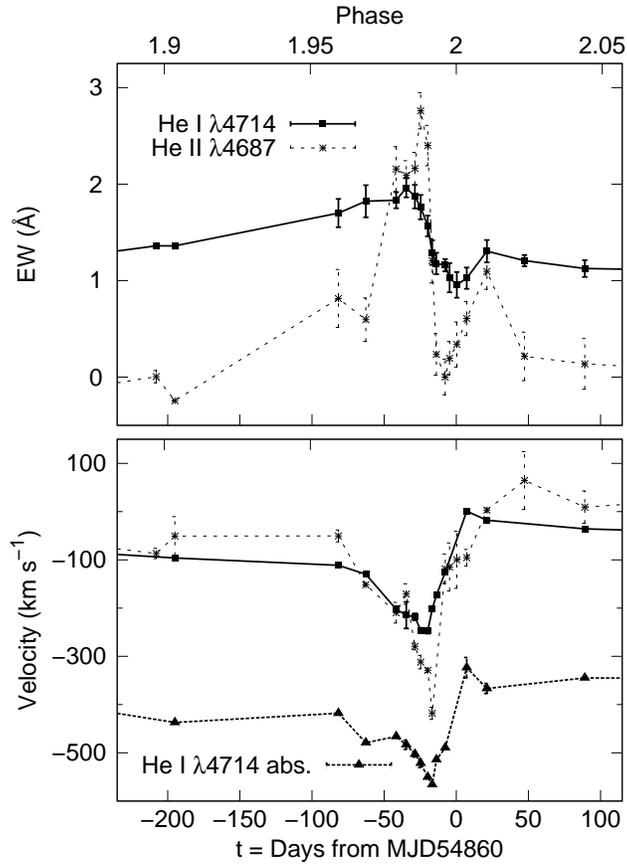}
\caption{Equivalent width and radial velocity of the \ion{He}{1} $\lambda$4714 emission (filled squares) on the star during the 2009 event observed with {\it Gemini\/} GMOS. The radial velocity of the absorption line is also shown (filled triangles). The crosses show the values for \ion{He}{2} $\lambda$4687 emission; the correlation is obvious. \label{fig:fig6}}
\end{figure}

\begin{figure}[!ht]  
\epsscale{0.75}
\plotone{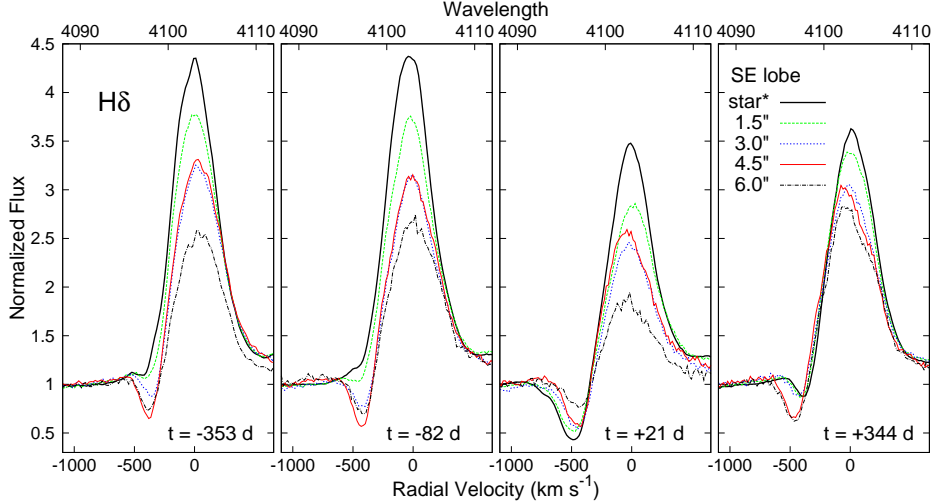}
\caption{H$\delta$ in tracings across the SE lobe in observations 
at $t = -353$ d (2008-02-11, phase 1.83), $t = -82$ d (2008-11-08, phase 1.96), 
$t = +21$ d (2009-02-19, phase 2.01), and $t = +344$ d (2010-01-08, phase 2.17).
The key indicates the positional offset along the spectrograph slit, SE 
from the central source. 
Spectra are shifted to compensate for their ${\Delta}V$ ``moving mirror''  
redshifts. The red curve (4\farcs5) corresponds to a spectrum which 
is close to position FOS4.  \label{fig:fig7} }
\end{figure}

\begin{figure}[!ht]   
\epsscale{0.75}
\plotone{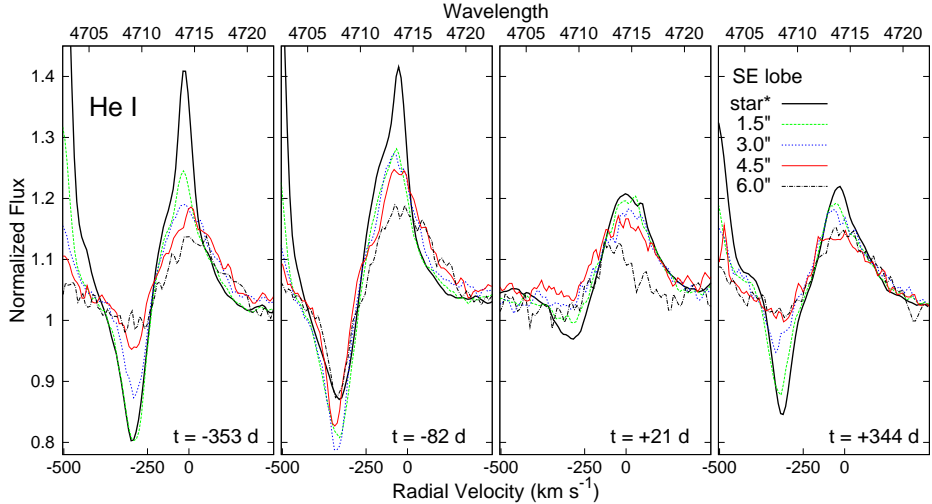}
\caption{Tracings across the SE lobe showing \ion{He}{1} $\lambda$4714 emission and absorption at the same times as Figure \ref{fig:fig7}. Velocities are corrected for the ``moving mirror'' redshift ${\Delta}V$. The red curve 
(4\farcs5) refers to position FOS4. Note that the large blueshift of the \ion{He}{1} emission and absorption lines during the events, discussed in Section 5,  cannot be observed in this Figure since it occurs between $t \approx -70$ d and $t \approx 0$ d not sampled here.  The feature at the left edge is 
[\ion{Fe}{3}] $\lambda$4703, which disappears during each spectroscopic 
event \citep{2010ApJ...710..729M}.\label{fig:fig8} }   
\end{figure}

\begin{figure}[!ht]   
\epsscale{0.5}
\plotone{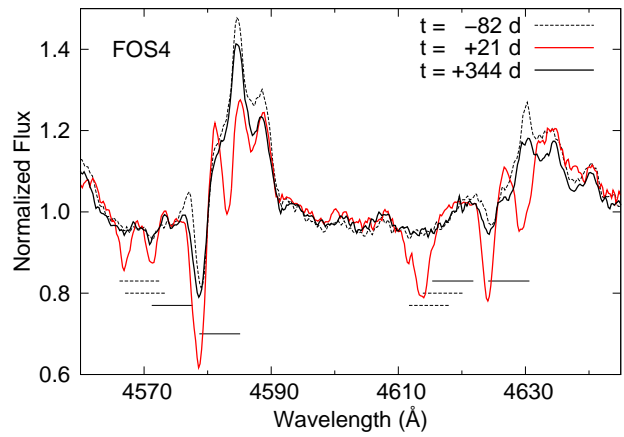}
\caption{Absorption components of \ion{Fe}{2} and similar ions at FOS4  
deepened for a few months during the 2009 event (solid red curve). They can be identified with absorption components at  $-$ 400--450 km s$^{-1}$ of \ion{Fe}{2} lines (solid black bars) and \ion{Cr}{2}, \ion{Mg}{1}, and \ion{Ti}{2} lines (dotted bars).\label{fig:fig9}}
\end{figure}

The semimajor axis $a$ is non-controversial, 16--19 AU for a total 
mass in the likely range 130--220 $M_\odot$.  Rough limits on the 
orbital eccentricity $\epsilon$ are usually based on the duration of 
a spectroscopic event.  Most likely the main parts of an event  
occur at orbital longitudes within $\pm 90\degree$ of periastron, 
i.e., while the two stars are separated less than twice the periastron 
distance.  Since the time required to move from $-90\degree$ to $+90\degree$ 
is ${\Delta}t_{180} \approx (1200 \, \mathrm{d}) \times (1 - \epsilon)^{3/2}$,  
eccentricities between 0.80 and 0.93 seem reasonably consistent with 
Figure \ref{fig:fig4}.  Most likely $0.84 < \epsilon < 0.91$,  
${\Delta}t_{180} \sim$ 30 to 80 days, and the periastron separation is 
between 1.5 and 2.8 AU.  If the first and second $\lambda$4687 episodes 
in 2009 occurred before and after periastron, the most likely time of 
periastron was close to $t \approx 0$.

The orbit orientation is more difficult to guess.  Very probably 
the orbit plane is close to the Homunculus midplane, with inclination 
$i \approx 45\degree$ \citep{2001AJ....121.1569D};  
any substantially different inclination seems unlikely in view 
of tidal friction near periastron, and would be outside the scope 
of our discussion here.   But what orientation does the orbit have 
within that plane?  Most authors quote $\omega$, the longitude of 
periastron as defined in textbooks.  For $\eta$ Car, $\omega = 270\degree$ 
would indicate that the secondary star is on the far side of the primary 
at periastron, while $\omega = 180\degree$ represents an orbit whose 
major axis is perpendicular to our line of sight.  
\citet{2008MNRAS.388L..39O}, \citet{2009MNRAS.394.1758P}, and    
\citet{2010A&A...517A...9G} all favored $\omega \approx 240\degree$ 
based on X-rays and other data, but these are not strictly 
independent estimates, since they shared a number of insecure 
assumptions.   One symptom of uncertainty is that 
\citet{2001ASPC..242...53I} found $i \approx 200\degree$    
by similar reasoning.  (The difference between 200\degree\ and 240\degree\ 
is substantial, because the latter value places the secondary star 
almost behind the primary at periastron, while 200\degree\ gives 
a more ``sideways'' view.)   Ishibashi's analysis was far simpler, 
but the other, more elaborate analyses required more assumptions.  
Moreover, \citet{2009MNRAS.397.1426K} used the X-rays to find an  
orientation much different from those cited above.  Many neglected 
effects can have major consequences on the wind simulations. For example, 
variable inhomogeneities commonly exist in stellar winds, may be 
especially crucial for $\eta$ Car's X-ray luminosity, and are 
extremely difficult to model.  Therefore $\omega$ is in fact 
quite uncertain.  

Another datum may be relevant. \citet{2010ApJ...710..729M} found conspicuous brightness maxima 
in the high-excitation [\ion{Ne}{3}] and [\ion{Fe}{3}] emission 
along our line of sight to the star, around phase 0.95 
or $t \sim -110$ d in 2003 -- i.e., considerably before the 
event had begun.  What is special about that part of the orbit?   
Considering the nature of those emission lines (see Mehner et al.),   
it probably signalled the time when our line of sight passes 
through the rarified secondary wind -- which corresponds to 
the time when the secondary star passed through the projection 
of our line of sight onto the orbit plane. If $\omega = 240\degree$   
and $\epsilon \lesssim 0.90$, this should have occurred at 
least 160 days before periastron.  A few calculations show two 
alternatives for $t \approx -110$ d:  $\omega \approx 240\degree$ and 
$\epsilon \approx 0.94$, which seems excessively eccentric, 
or $\omega$ in the range 220\degree--232\degree\ if 
$0.85 \lesssim \epsilon \lesssim 0.90$. 
At any rate {\it the orientation of the orbit is not well 
established.\/}\footnote{   
     Incidentally, the high-excitation precursor peak should recur 
     in April-May 2014, and, if our interpretation is valid, it 
     should be broader and less conspicuous than in 2003.  Reason: 
     The decreased primary wind should have broadened the opening 
     angle of the wind-wind ``shock cone.'' \label{foot:foot9}}     

\paragraph{Did an eclipse play any role in the spectroscopic event?}  
For the most popular orbit parameters (see above), the 
secondary star should have passed behind the far side of the primary     
wind in the weeks following periastron.  The intervening wind might conceivably have been opaque in 2003 
but not in 2009.   For a credible decrease in density between those 
times, however, the wind should not have been {\it very\/} opaque 
in 2003;  the eclipse ingress and egress should not have been 
abrupt.   See remarks in \citet{2006ApJ...640..474M}.

{\it Acknowledgements}
We thank the staff and observers of
the Gemini-South Observatory in La Serena for their help in preparing and conducting the observations, and Beth Perriello at STScI for assistance with {\it HST\/} observing plans.


\appendix

\newcommand{\appsection}[1]{\let\oldthesection\thesection
  \renewcommand{\thesection}{Appendix \oldthesection}
  \section{#1}\let\thesection\oldthesection}

\appsection{Concerning ``Phase'' and the 5.5-year Period}   
\label{app}

We know from experience that varying definitions of ``phase'' in 
$\eta$ Car's 5.5-year cycle have caused confusion.  There have been 
at least three difficulties, some of them matters of principle:  
  \begin{enumerate} 
  \item Most authors specified their zero points $t_0$ to coincide with 
    various observed phenomena, e.g.\ the disappearance of some spectral 
    feature.   This policy gives an impression that the definition 
    is based on known physical effects, correlated with theory. 
    In fact, the spectroscopic events are too poorly understood to 
    identify any observable quantity with a basic physical or geometric 
    parameter in the binary system.  
  \item This style of reckoning gives a false impression of high 
     precision and future repeatability.  In fact, successive events 
     are known to differ from each other.  Given the continuing 
     changes in $\eta$ Car 
     (see refs.\ in Section 1), any observable quantity may shift its phase 
     in the cycle. Terms such as ``ephemeris'' can be very misleading 
     in circumstances like these. 
  \item The adopted periods and zero points have varied from paper 
     to paper!  We have identified at least seven different phase 
     definitions of this type in the literature, with periods spanning 
     a range of 10 days and zero points ranging across 20 days.  
     Authors have repeatedly changed their definitions.
  \end{enumerate} 

Fortunately one simple definition has remained 
constant, and is greatly preferable in terms of logic and procedure.  
For the $\eta$ Car HST Treasury Program Archive 
(http://etacar.umn.edu/),  
``phase'' was intentionally defined in a calendar-based rather than a 
phenomenon-based way.  This choice was made specifically to avoid 
the pitfalls listed above.  Its period and zero point are 
2023.0 days  and MJD 50814.0 = J1998.000 exactly.  
Phase 2.000 occurred at MJD 50860.0 = J2009.077.  The integer 
quantities help to minimize calculative errors, and, more important,  
they discourage any impression that $t = 0$ represents some critical 
time with respect to physics.  (Periastron is thought to occur 
fairly close to $t = 0$, but this categorically plays no 
role in the definition.)

This system has been in use since 2003 without alteration;  its adopted 
period of 2023.0 days continues to agree with the best measurements within 
less than 1$\sigma$ \citep{2008MNRAS.384.1649D,2010NewA...15..108F}; 
and it is used in the Treasury Program archive, which is the largest 
easily-accessible source of data on $\eta$ Car (reduced data  of all {\it HST\/} STIS, {\it VLT\/} UVES, and {\it Gemini\/} GMOS observations are available).  {\it Hence there is no 
reason to substitute any later, arbitrary system.\/}    In order to 
minimize present and future confusion, and for the other reasons 
noted above, it is the obvious standard.

Timing measurements are valuable because they may indicate changes between 
successive events.  One particular operational procedure achieves the best 
precision and reproducibility, as follows.  Various observable quantitities 
-- e.g., photometry at most wavelengths -- attain brief maxima 
$Q_\mathrm{max}$ at a particular stage in a spectroscopic event, and then 
briskly fall to their local minima $Q_\mathrm{min}$ (or vice-versa).  The 
times of maximum and minimum are imprecise, but for some observables 
the time of midpoint, when $Q$ is the average of maximum and minimum, 
is more precise than anything else in the spectroscopic event!  This 
is true mainly because the descent is very rapid at that time.  If 
$Q(t)$ is fairly smooth  and if enough data points are available, 
then the most robust measurement protocol is as follows.  
  \begin{enumerate}  
  \item  Estimate the value of $Q_\mathrm{max}$ but ignore its 
     time $t(Q_\mathrm{max})$ which may be ill-defined.  
     A local quadratic fit may be appropriate, but fortunately 
     there is no need for very high precision in $Q_\mathrm{max}$ 
     (see below).  
  \item Do the same for $Q_\mathrm{min}$, and calculate the 
     midpoint $Q_{\mathrm{m}} = (Q_\mathrm{min} + Q_\mathrm{max})/2$.   
  \item Then estimate the midpoint time $t_\mathrm{m} = t(Q_\mathrm{m})$, 
     {\it based only on the data points that are nearest to it.}    
     If many data points are available, for instance, one might 
     use only those in the range $Q_\mathrm{m} \pm {\Delta}Q$ 
     where ${\Delta}Q \approx (Q_\mathrm{max} - Q_\mathrm{min})/4$.  
     In favorable circumstances either a linear or a cubic fit 
     gives practically the same result.    
   \end{enumerate} 
In effect this is a specialized form of interpolation.  Each step employs 
only a local subset of the data, so local irregularities in behavior 
around the maximum and minimum do not 
affect $t_\mathrm{m}$ much.  In the case of $\eta$ Car's events, 
$t_\mathrm{m}$ is only weakly sensitive to the estimated values  
$Q_\mathrm{max}$ and $Q_\mathrm{min}$.  This is true because, for 
most observables, the curve $Q(t)$  
is approximately antisymmetric near the midpoint and the rate 
of descent is quite rapid there.  Note that theoretical significance 
plays no role in the empirical reasoning.  For an especially successful 
example, $\eta$ Car's `J' magnitudes in 2003 reported by 
\citet{2004MNRAS.352..447W}, the rms formal error in $t_\mathrm{m}$ 
is only $\pm 0.4$ day.  We employed the same method for some 
details in  Section \ref{sec:episode1} above.

\citet{2010NewA...15..108F} used a method that is more general but 
is not as well suited to $\eta$ Car in particular.  In effect, their 
procedure mixes data in a broader time interval, thereby allowing a 
stronger dependence on behavior details around $Q_\mathrm{max}$ and 
$Q_\mathrm{min}$.   \citealt{2008MNRAS.384.1649D} defined a phase 
zero point based on an extrapolation, which of course is inherently  
far less robust than interpolation.

\citet{2009A&A...493.1093F,2010NewA...15..108F} report 
high-quality ground-based  photometry since 2003, but it is extremely 
unfortunate that the earlier JHK program described by 
\citet{2001MNRAS.322..741F} and 
\citet{2004MNRAS.352..447W} was forced to close before the 2009.0 
event.  Their results give precise estimates of a time related to 
the 2003.5 event, and their earlier data give fairly good results 
for the 1998.0 event (see below).  Here we estimate the 2003--2009 
inter-event time interval from our {\it HST\/} UV photometry 
(Section \ref{sec:phot}), because these data represent the stellar 
wind with little contamination by ejecta at $r \gtrsim 0.15\arcsec$. 
Unfortunately the temporal sampling is sparse compared to 
ground-based photometry.  

Table \ref{tab:table4} lists the results for $t_\mathrm{m}$ in three events and 
five wavebands.  Here we use JHK photometry by Feast et al.\ and Whitelock 
et al.\ cited above, and our UV photometry.   JHK wavelengths are pertinent 
because light reflected by dust in the ejecta is less than for UBVRI.  
The quoted uncertainties are semi-formal rms standard errors based 
on observed scatter or deviation of the data points and other details;  
these appear to be realistic, judging from the scatter in the three 
periods based independently on J, H, and K.  Note 
that $t_\mathrm{m}$ for a particular event may depend on 
wavelength;  it is the {\it difference\/} between events that we 
are concerned with here.   The JHK data indicate an interval 
2021.5 $\pm$ 0.9 d between the 1998.0 and 2003.5 events;  this compares 
with 2022.7 $\pm$ 1.3 d found by \citet{2008MNRAS.384.1649D} with less 
suitable methods and a much larger set of data, and 2022.8 $\pm$ 0.5 d 
found by \citet{2010NewA...15..108F} from ground-based photometry 
in 2003--2009.  The {\it HST\/} 330 nm value is obviously consistent 
with these,  but the 250 nm result appears discordant.  This may be a real 
effect;  the 250 nm wavelength region physically differs from 
the others because it represents the \ion{Fe}{2} forest 
(Section \ref{sec:phot}).  Unfortunately there is no assurance that 
UV photometry will be possible during the next, 2014.6 event.




\newpage

\begin{deluxetable}{llcc}   
  \tablewidth{0pt}
  \tabletypesize{\scriptsize}
  \tablecaption{{\it HST\/} Photometry Results\tablenotemark{a}\label{tab:table1}}
  \tablecolumns{4}
  \tablehead{
    &
    &
    \colhead{Flux Density}&
    \colhead{STMAG\tablenotemark{b}}\\
    \colhead{MJD}&
    \colhead{Year}&
    \colhead{(10$^{-11}$ erg cm$^{-2}$ s$^{-1}$ \mbox{\AA}$^{-1}$)}&
    \colhead{(mag)}
  }
  \startdata
  \multicolumn{4}{c}{WFPC2 PC/F255W Filter}\\
  54717.1&2008.69&1.074&6.323\\
  54787.8&2008.88&1.126$\pm$0.013&6.271$\pm$0.013\\
  54807.4&2008.93&1.099$\pm$0.035&6.299$\pm$0.035\\
  54834.0&2009.01&1.189$\pm$0.004&6.212$\pm$0.004\\
  54838.4&2009.02&0.967$\pm$0.014&6.437$\pm$0.016\\
  54841.5&2009.03&0.806$\pm$0.006&6.635$\pm$0.008\\
  54845.5&2009.04&0.598$\pm$0.017&6.959$\pm$0.031\\
  54854.5&2009.06&0.451$\pm$0.003&7.265$\pm$0.008\\
  54872.4&2009.11&0.437$\pm$0.019&7.298$\pm$0.048\\
  54949.2&2009.32&0.937$\pm$0.011&6.471$\pm$0.013\\        
  \tableline
  \multicolumn{4}{c}{WFPC2 PC/F336W Filter}\\
  54717.1&2008.69&1.108$\pm$0.035&6.289$\pm$0.034\\
  54787.8&2008.88&1.114$\pm$0.041&6.283$\pm$0.040\\
  54807.4&2008.93&1.142$\pm$0.031&6.256$\pm$0.030\\
  54834.0&2009.01&1.351$\pm$0.005&6.073$\pm$0.004\\
  54838.4&2009.02&1.399$\pm$0.037&6.036$\pm$0.029\\
  54841.5&2009.03&1.388$\pm$0.023&6.045$\pm$0.018\\
  54845.5&2009.04&1.300$\pm$0.007&6.116$\pm$0.006\\
  54854.5&2009.06&1.004$\pm$0.020&6.396$\pm$0.022\\
  54872.4&2009.11&1.008$\pm$0.027&6.392$\pm$0.029\\
  54949.2&2009.32&1.599$\pm$0.009&5.890$\pm$0.006\\
    \tableline
  \multicolumn{4}{c}{STIS Synthetic/F250W Filter}\\
55062.0&2009.63&&6.12\\
55258.7&2010.17&&5.89\\
55428.3&2010.63&&5.65 \\
  \tableline
  \multicolumn{4}{c}{STIS Synthetic/F330W Filter}\\
55062.0&2009.63&&5.63\\
55258.7&2010.17&&5.55\\
55428.3&2010.63&&5.35\\
  \enddata
  \tablenotetext{a}{Earlier results can be found in \citep{2004AJ....127.2352M,2006AJ....132.2717M,2010AJ....139.2056M}.}
  \tablenotetext{b}{The STMAG photometric system is calibrated for direct comparison of fluxes with similar filters in different instruments.}
\end{deluxetable}

\begin{deluxetable}{lcccccc}    
\tabletypesize{\scriptsize} 
\tablecaption{{\it Gemini\/} GMOS Journal\tablenotemark{a}\label{tab:table2}}   
\tablewidth{0pt} 
\tablehead{ 
\colhead{} & 
\colhead{} &  
\colhead{}&
\colhead{}&
\colhead{}&
\colhead{Star\tablenotemark{c}} &
\colhead{FOS4\tablenotemark{d}}\\
\colhead{Name\tablenotemark{b}} & 
\colhead{Date} &  
\colhead{MJD}&
\colhead{Phase}&
\colhead{Cenwave}&
 \colhead{Exp. time\tablenotemark{e}} &
\colhead{Exp. time\tablenotemark{e}}\\
\colhead{} & 
\colhead{} &   
\colhead{} & 
\colhead{} &
\colhead{($\lambda$)} & 
\colhead{(s)} &
\colhead{(s)} 
 }   
\startdata
gH45	&	Jun 16, 2007	&	54267.1	&	1.707	&	4300	&	10 & --	\\
gH45	&	Jun 17, 2007	&	54269.0	&	1.708	&	4300	&	10 & --	\\
gH49	&	Jun 30, 2007	&	54281.0	&	1.714	&	5600	&	40 & 220 	\\
gI11	&	Feb 11, 2008	&	54507.4	&	1.826	&	4300	&	77 & 	377\\
gI11	&	Feb 13, 2008	&	54509.2	&	1.827	&	5600	&	4 & 	260 \\
gI50	&	Jul 5, 2008	&	54652.0	&	1.897	&	5600	&	40 & 260	\\
gI54	&	Jul 17, 2008	&	54665.0	&	1.904	&	4300	&	10 & 300	\\
gI85	&	Nov 8, 2008	&	54778.3	&	1.960	&	5200	&	40 & 453	\\
gI85	&	Nov 8, 2008	&	54778.3	&	1.960	&	4300	&	77 & 377	\\
gI90	&	Nov 27, 2008	&	54797.3	&	1.969	&	5200	&	40 & 450	\\
gI90	&	Nov 27, 2008	&	54797.3	&	1.969	&	4300	&	77 & 377	\\
gI96	&	Dec 18, 2008	&	54818.3	&	1.979	&	5200	&	40 & 450	\\
gI96	&	Dec 18, 2008	&	54818.4	&	1.979	&	4300	&	7	 & 377 \\
gI98	&	Dec 25, 2008	&	54825.3	&	1.983	&	5200	&	40 & 	450\\
gI98	&	Dec 25, 2008	&	54825.4	&	1.983	&	4300	&	77 & 377	\\
gI99	&	Dec 31, 2008	&	54831.3	&	1.986	&	5200	&	40 & 450	\\
gI99	&	Dec 31, 2008	&	54831.4	&	1.986	&	4300	&	10 & 	70\\
gJ01 &	Jan 4, 2009	&	54835.3	&	1.988	&	5200	&	45	 & 450 \\
gJ01 &	Jan 4, 2009	&	54835.3	&	1.988	&	4300	&	55 & 	377 \\
gJ02	&	Jan 9, 2009	&	54840.2	&	1.990	&	5200	&	45 & 450	\\
gJ02	&	Jan 9, 2009	&	54840.2	&	1.990	&	4300	&	33 & 377	\\
gJ03	&	Jan 12, 2009	&	54843.3	&	1.992	&	5200	&	45 & 	450\\
gJ03	&	Jan 12, 2009	&	54843.3	&	1.992	&	4300	&	6	 & 377 \\
gJ04	&	Jan 15, 2009	&	54846.2	&	1.993	&	5200	&	15 & 	450 \\
gJ04	&	Jan 15, 2009	&	54846.2	&	1.993	&	4300	&	33 & 377	\\
gJ05	&	Jan 21, 2009	&	54852.3	&	1.996	&	5200	&	4	 & 450 \\
gJ05	&	Jan 21, 2009	&	54852.3	&	1.996	&	4300	&	6	 & 300 \\
gJ06	&	Jan 24, 2009	&	54855.3	&	1.998	&	5200	&	4	 & 260 \\
gJ06	&	Jan 24, 2009	&	54855.4	&	1.998	&	4300	&	6	 & 300 \\
gJ07	&	Jan 29, 2009	&	54860.4	&	2.000	&	5200	&	9	 & 450 \\
gJ07	&	Jan 29, 2009	&	54860.4	&	2.000	&	4300	&	33 & 377	\\
gJ09	&	Feb 5, 2009	&	54867.2	&	2.004	&	5200	&	15 & 450	\\
gJ09	&	Feb 5, 2009	&	54867.3	&	2.004	&	4300	&	6	 & 300 \\
gJ13	&	Feb 19, 2009	&	54881.2	&	2.010	&	5200	&	4 & 450	\\
gJ13	&	Feb 19, 2009	&	54881.3	&	2.011	&	4300	&	6 & 	300 \\
gJ20	&	Mar 17, 2009	&	54907.3	&	2.023	&	5200	&	4 & --	\\
gJ20	&	Mar 17, 2009	&	54907.3	&	2.023	&	4300	&	6 & --	\\
gJ32	&	Apr 28, 2009	&	54949.1	&	2.044	&	5200	&	4 & 	260\\
gJ32	&	Apr 28, 2009	&	54949.1	&	2.044	&	4300	&	6 & 	300\\
gJ56	&	Jul 23, 2009	&	55036.0	&	2.087	&	5200	&	4	 & 260  \\
gJ56	&	Jul 24, 2009	&	55036.0	&	2.087	&	4300	&	77 &  300	\\
gK02	&	Jan 8, 2010	&	55204.32	&	2.170	&	5200	&	4	 & --\\
gK02	&	Jan 8, 2010	&	55204.34	&	2.170	&	4300	&	4	 & --\\
gK05	&	Jan 20, 2010	&	55216.29	&	2.176	&	5200	&	-- & 260	\\
\enddata 
\tablenotetext{a}{PA = 160{\degree}, Grating: B1200\_G5321.} 
\tablenotetext{b}{As listed in http://etacar.umn.edu/.} 
\tablenotetext{c}{Include slit positions less than $\pm$0\farcs375 from the star at $\lambda$4687.} 
\tablenotetext{d}{Include slit positions -1 to -3{\arcsec} from the star at $\lambda$4687.} 
\tablenotetext{e}{Several exposures were taken on each data, combined exposure times are listed.} 
\end{deluxetable}

\begin{deluxetable}{lccccc}    
\tabletypesize{\scriptsize} 
\tablecaption{Equivalent width and radial velocity of \ion{He}{2} $\lambda$4687 in {\it Gemini\/} GMOS data on the star\label{tab:table3}} 
\tablewidth{0pt} 
\tablehead{ 
   \colhead{Name\tablenotemark{a}} &
   \colhead{Date} &
   \colhead{MJD}   &
   \colhead{Phase} &
   \colhead{EW} &
   \colhead{$V_\textnormal{rad}^b$} \\ 
   \colhead{} &
   \colhead{(UT)} &
   \colhead{}   &
   \colhead{} &
   \colhead{(\AA)} &
   \colhead{(km s$^{-1}$)} 
 }   
\startdata   
    gH45	&	2007 Jun 16	&	54268.0	&	1.707	&	-0.25	 $\pm$ 	0.02	&	15.09 $\pm$ 6.11 \\
    gH49	&	2007 Jun 30	&	54281.0	&	1.714	&	-0.19	 $\pm$ 	0.01	&	-39.60 $\pm$ 46.88	\\
	gI11	&	2008 Feb 11	&	54507.4	&	1.826	&	-0.34	 $\pm$ 	0.03	&	-35.39 $\pm$ 54.12	\\
	gI50	&	2008 Jul 05	&	54652.0	&	1.897	&	0.01	 $\pm$ 	0.06	&	-86.80 $\pm$ 10.60	\\
	gI54	&	2008 Jul 17	&	54665.0	&	1.904	&	-0.25	 $\pm$ 	0.01	&	-50.83 $\pm$ 40.44	\\
	gI85	&	2008 Nov 08	&	54778.3	&	1.960	&	0.82	 $\pm$ 	0.30	&	-50.46 $\pm$ 12.29	\\
	gI90	&	2008 Nov 27	&	54797.3	&	1.969	&	0.60	 $\pm$ 	0.23	&	-151.68 $\pm$ 2.73	\\
	gI96	&	2008 Dec 18	&	54818.3	&	1.979	&	2.16	 $\pm$ 	0.23	&	-209.59 $\pm$ 21.47	\\
	gI98 &	2008 Dec 25	&	54825.3	&	1.983	&	2.09	 $\pm$ 	0.15	&	-170.93 $\pm$ 21.29	\\
	gI99	&	2008 Dec 31	&	54831.3	&	1.986	&	2.16	 $\pm$ 	0.17	&	-279.30 $\pm$ 7.06	\\
	gJ01	&	2009 Jan 04	&	54835.3	&	1.988	&	2.76	 $\pm$ 	0.19	&	-311.94 $\pm$ 13.91	\\
	gJ02	&	2009 Jan 09	&	54840.2	&	1.990	&	2.40	 $\pm$ 	0.21	&	-328.94 $\pm$ 4.58	\\
	gJ03	&	2009 Jan 12	&	54843.3	&	1.992	&	1.21	 $\pm$ 	0.23	&	-418.11 $\pm$ 12.49	\\
	gJ04	&	2009 Jan 15	&	54846.2	&	1.993	&	0.24	 $\pm$ 	0.22	&	-- 	\\
	gJ05 &	2009 Jan 21	&	54852.3	&	1.996	&	0.00	 $\pm$ 	0.18	&	-118.72 $\pm$ 30.72	\\
	gJ06	&	2009 Jan 24	&	54855.3	&	1.998	&	0.20	 $\pm$ 	0.17	&	-114.85 $\pm$ 49.71	\\
	gJ07	&	2009 Jan 29	&	54860.4	&	2.000	&	0.34	 $\pm$ 	0.23	&	-99.95 $\pm$ 59.14	\\
	gJ09	&	2009 Feb 05	&	54867.2	&	2.004	&	0.61	 $\pm$ 	0.18	&	-95.55 $\pm$ 17.06\\
	gJ13	&	2009 Feb 19	&	54881.2	&	2.011	&	1.10	 $\pm$ 	0.19	&	3.08 $\pm$ 3.97		\\
	gJ20	&	2009 Mar 17	&	54907.3	&	2.023	&	0.22	 $\pm$ 	0.25	&	64.57 $\pm$ 59.78	\\
	gJ32	&	2009 Apr 28	&	54949.1	&	2.044	&	0.14	 $\pm$ 	0.26	&	9.28 $\pm$ 33.48	\\
	gJ56	&	2009 Jul 24	&	55036.0	&	2.087	&	0.06	 $\pm$ 	0.30	&	27.39 $\pm$ 19.36	\\
	gK02 &	2010 Jan 08	&	55204.3	&	2.170	&	0.08 $\pm$ 0.19	&	-56.88 $\pm$ 7.80 \\
	\enddata 
	\tablenotetext{a}{As listed on the Eta Carinae Treasury Project site at http://etacar.umn.edu/.} 
\end{deluxetable}

\begin{deluxetable}{lcccc}    
\tabletypesize{\scriptsize} 
\tablecaption{Photometric $t_{phot}$ MJD's for $\eta$ Car's 1998.0, 2003.5, and 2009.0 spectroscopic events \label{tab:table4}}   
\tablewidth{0pt} 
\tablehead{ 
\colhead{} & 
\colhead{1998.0} &  
\colhead{2003.5}&
\colhead{2009.0}&
\colhead{Differenz}\\
\colhead{Waveband} & 
\colhead{$t_{phot}$} &   
\colhead{$t_{phot}$} & 
\colhead{$t_{phot}$} &
\colhead{$\Delta t_{phot}$} 
 }   
\startdata       
K\tablenotemark{a}   &  50802.6 $\pm$ 1.8  &  52824.7 $\pm$ 0.5  &  ---  &  2022.1 $\pm$ 1.9 d  \\   
H\tablenotemark{a}   &  50803.3 $\pm$ 1.3  &  52825.9 $\pm$ 0.4  &  ---  &  2022.6 $\pm$ 1.4 d  \\   
J\tablenotemark{a}    &  50807.0 $\pm$ 1.5  &  52826.8 $\pm$ 0.4  &  ---  &  2019.8 $\pm$ 1.6 d  \\   
330 nm &  ---  &  52827.8 $\pm$ 1.7  &  54850.0 $\pm$ 1.7  &  2022.2 $\pm$ 2.4 d  \\
250 nm &  ---  &  52826.1 $\pm$ 1.1  &  54842.7 $\pm$ 0.6  &  2016.6 $\pm$ 1.3 d  \\ 
\enddata 
\tablenotetext{a}{\citet{2001MNRAS.322..741F,2004MNRAS.352..447W}.}
\end{deluxetable}

\end{document}